\documentclass[draftcls,onecolumn,11pt]{IEEEtran}
\usepackage{amsmath,graphicx}

\usepackage[T1]{fontenc}
\usepackage[utf8]{inputenc}
\usepackage{amsmath,amssymb,amsfonts,mathrsfs,bm}
\usepackage{mathtools}
\usepackage{amsthm}
\usepackage{scalerel}
\usepackage{nicefrac}
\usepackage[shortlabels]{enumitem}
\usepackage{graphicx}
\usepackage{epstopdf}
\DeclareGraphicsExtensions{.eps,.png,.jpg,.pdf}

\usepackage{url}
\usepackage{colortbl}
\usepackage{booktabs}
\usepackage{multirow}
\usepackage[table,dvipsnames]{xcolor}
\usepackage[normalem]{ulem}
\usepackage{xparse}
\usepackage{calc}
\usepackage{etoolbox}

\makeatletter
\@ifpackageloaded{natbib}{
	\relax
}{
	\usepackage{cite}
}
\makeatother

\usepackage{array}
\newcolumntype{L}[1]{>{\raggedright\let\newline\\\arraybackslash\hspace{0pt}}m{#1}}
\newcolumntype{C}[1]{>{\centering\let\newline\\\arraybackslash\hspace{0pt}}m{#1}}
\newcolumntype{R}[1]{>{\raggedleft\let\newline\\\arraybackslash\hspace{0pt}}m{#1}}

\makeatletter
\let\MYcaption\@makecaption
\makeatother
\usepackage[font=footnotesize]{subcaption}
\makeatletter
\let\@makecaption\MYcaption
\makeatother

\usepackage{glossaries}
\makeatletter
\sfcode`\.1006

\let\oldgls\gls
\let\oldglspl\glspl

\newcommand\fussy@ifnextchar[3]{%
	\let\reserved@d=#1%
	\def\reserved@a{#2}%
	\def\reserved@b{#3}%
	\futurelet\@let@token\fussy@ifnch}
\def\fussy@ifnch{%
	\ifx\@let@token\reserved@d
		\let\reserved@c\reserved@a
	\else
		\let\reserved@c\reserved@b
	\fi
	\reserved@c}

\renewcommand{\gls}[1]{%
\oldgls{#1}\fussy@ifnextchar.{\@checkperiod}{\@}}
\renewcommand{\glspl}[1]{%
\oldglspl{#1}\fussy@ifnextchar.{\@checkperiod}{\@}}

\newcommand{\@checkperiod}[1]{%
	\ifnum\sfcode`\.=\spacefactor\else#1\fi
}
\makeatother

\newacronym{wrt}{w.r.t.}{with respect to}
\newacronym{RHS}{R.H.S.}{right-hand side}
\newacronym{LHS}{L.H.S.}{left-hand side}
\newacronym{iid}{i.i.d.}{independent and identically distributed}

\usepackage{float}

\ifx\notloadhyperref\undefined
	\ifx\loadbibentry\undefined
		\usepackage[hidelinks,hypertexnames=false]{hyperref}
	\else
		\usepackage{bibentry}
		\makeatletter\let\saved@bibitem\@bibitem\makeatother
		\usepackage[hidelinks,hypertexnames=false]{hyperref}
		\makeatletter\let\@bibitem\saved@bibitem\makeatother
	\fi
\else
	\ifx\loadbibentry\undefined
		\relax
	\else
		\usepackage{bibentry}
	\fi
\fi

\usepackage[capitalize]{cleveref}
\crefname{equation}{}{}
\Crefname{equation}{}{}
\crefname{claim}{claim}{claims}
\crefname{step}{step}{steps}
\crefname{line}{line}{lines}
\crefname{condition}{condition}{conditions}
\crefname{dmath}{}{}
\crefname{dseries}{}{}
\crefname{dgroup}{}{}

\crefname{Problem}{Problem}{Problems}
\crefformat{Problem}{Problem~(#2#1#3)}
\crefrangeformat{Problem}{Problems~(#3#1#4) to~(#5#2#6)}

\crefname{Theorem}{Theorem}{Theorems}
\crefname{Corollary}{Corollary}{Corollaries}
\crefname{Proposition}{Proposition}{Propositions}
\crefname{Lemma}{Lemma}{Lemmas}
\crefname{Definition}{Definition}{Definitions}
\crefname{Example}{Example}{Examples}
\crefname{Assumption}{Assumption}{Assumptions}
\crefname{Remark}{Remark}{Remarks}
\crefname{Rem}{Remark}{Remarks}
\crefname{remarks}{Remarks}{Remarks}
\crefname{Appendix}{Appendix}{Appendices}
\crefname{Supplement}{Supplement}{Supplements}
\crefname{Exercise}{Exercise}{Exercises}
\crefname{Theorem_A}{Theorem}{Theorems}
\crefname{Corollary_A}{Corollary}{Corollaries}
\crefname{Proposition_A}{Proposition}{Propositions}
\crefname{Lemma_A}{Lemma}{Lemmas}
\crefname{Definition_A}{Definition}{Definitions}

\usepackage{crossreftools}
\ifx\notloadhyperref\undefined
\else
	\relax
\fi

\usepackage{algorithm,algorithmic}

\ifx\loadbreqn\undefined
	\relax
\else
	\usepackage{breqn}
\fi

\makeatletter
\def\cleartheorem#1{%
    \expandafter\let\csname#1\endcsname\relax
    \expandafter\let\csname c@#1\endcsname\relax
}
\def\clearthms#1{ \@for\tname:=#1\do{\cleartheorem\tname} }
\makeatother

\ifx\renewtheorem\undefined
	\ifx\useTheoremCounter\undefined
		\newtheorem{Theorem}{Theorem}
		\newtheorem{Corollary}{Corollary}
		\newtheorem{Proposition}{Proposition}
		\newtheorem{Lemma}{Lemma}
	\else
		\newtheorem{Theorem}{Theorem}
		\newtheorem{Corollary}[Theorem]{Corollary}
		\newtheorem{Proposition}[Theorem]{Proposition}
	\fi

	\newtheorem{Definition}{Definition}
	
	\newtheorem{Remark}{Remark}

\fi

\theoremstyle{remark}

\theoremstyle{plain}

\newcommand{\qednew}{\nobreak \ifvmode \relax \else
		\ifdim\lastskip<1.5em \hskip-\lastskip
			\hskip1.5em plus0em minus0.5em \fi \nobreak
		\vrule height0.75em width0.5em depth0.25em\fi}



\newcommand{\nn}{\nonumber\\ }

\NewDocumentCommand{\movedownsub}{e{^_}}{%
	\IfNoValueTF{#1}{%
		\IfNoValueF{#2}{^{}}
	}{%
		^{#1}
	}%
	\IfNoValueF{#2}{_{#2}}
}

\let\latexchi\chi
\RenewDocumentCommand{\chi}{}{\latexchi\movedownsub}

\newcommand{\Real}{\mathbb{R}}

\newcommand{\calA}{\mathcal{A}}
\newcommand{\calB}{\mathcal{B}}

\newcommand{\calD}{\mathcal{D}}

\newcommand{\calF}{\mathcal{F}}

\newcommand{\calH}{\mathcal{H}}

\newcommand{\calL}{\mathcal{L}}
\newcommand{\calM}{\mathcal{M}}
\newcommand{\calN}{\mathcal{N}}

\newcommand{\calR}{\mathcal{R}}
\newcommand{\calS}{\mathcal{S}}

\newcommand{\calW}{\mathcal{W}}
\newcommand{\calX}{\mathcal{X}}
\newcommand{\calY}{\mathcal{Y}}
\newcommand{\calZ}{\mathcal{Z}}

\newcommand{\bA}{\mathbf{A}}

\newcommand{\bI}{\mathbf{I}}

\newcommand{\bx}{\mathbf{x}}

\newcommand{\bbP}{\mathbb{P}}
\newcommand{\bbQ}{\mathbb{Q}}
\newcommand{\bbR}{\mathbb{R}}

\newcommand{\scP}{\mathscr{P}}

\DeclareSymbolFont{bsfletters}{OT1}{cmss}{bx}{n}
\DeclareSymbolFont{ssfletters}{OT1}{cmss}{m}{n}
\DeclareMathSymbol{\bsfGamma}{0}{bsfletters}{'000}
\DeclareMathSymbol{\ssfGamma}{0}{ssfletters}{'000}
\DeclareMathSymbol{\bsfDelta}{0}{bsfletters}{'001}
\DeclareMathSymbol{\ssfDelta}{0}{ssfletters}{'001}
\DeclareMathSymbol{\bsfTheta}{0}{bsfletters}{'002}
\DeclareMathSymbol{\ssfTheta}{0}{ssfletters}{'002}
\DeclareMathSymbol{\bsfLambda}{0}{bsfletters}{'003}
\DeclareMathSymbol{\ssfLambda}{0}{ssfletters}{'003}
\DeclareMathSymbol{\bsfXi}{0}{bsfletters}{'004}
\DeclareMathSymbol{\ssfXi}{0}{ssfletters}{'004}
\DeclareMathSymbol{\bsfPi}{0}{bsfletters}{'005}
\DeclareMathSymbol{\ssfPi}{0}{ssfletters}{'005}
\DeclareMathSymbol{\bsfSigma}{0}{bsfletters}{'006}
\DeclareMathSymbol{\ssfSigma}{0}{ssfletters}{'006}
\DeclareMathSymbol{\bsfUpsilon}{0}{bsfletters}{'007}
\DeclareMathSymbol{\ssfUpsilon}{0}{ssfletters}{'007}
\DeclareMathSymbol{\bsfPhi}{0}{bsfletters}{'010}
\DeclareMathSymbol{\ssfPhi}{0}{ssfletters}{'010}
\DeclareMathSymbol{\bsfPsi}{0}{bsfletters}{'011}
\DeclareMathSymbol{\ssfPsi}{0}{ssfletters}{'011}
\DeclareMathSymbol{\bsfOmega}{0}{bsfletters}{'012}
\DeclareMathSymbol{\ssfOmega}{0}{ssfletters}{'012}

\newcommand{\btheta}{\bm{\theta}}
\newcommand{\bmu}{\bm{\mu}}
\newcommand{\bnu}{\bm{\nu}}
\newcommand{\btau}{\bm{\tau}}

\newcommand{\bsigma}{\bm{\sigma}}

\newcommand{\bphi}{\bm{\phi}}

\makeatletter
\newcommand*\rel@kern[1]{\kern#1\dimexpr\macc@kerna}
\newcommand*\widebar[1]{%
  \begingroup
  \def\mathaccent##1##2{%
    \rel@kern{0.8}%
    \overline{\rel@kern{-0.8}\macc@nucleus\rel@kern{0.2}}%
    \rel@kern{-0.2}%
  }%
  \macc@depth\@ne
  \let\math@bgroup\@empty \let\math@egroup\macc@set@skewchar
  \mathsurround\z@ \frozen@everymath{\mathgroup\macc@group\relax}%
  \macc@set@skewchar\relax
  \let\mathaccentV\macc@nested@a
  \macc@nested@a\relax111{#1}%
  \endgroup
}
\makeatother

\DeclareMathOperator*{\argmax}{arg\,max}

\DeclareMathOperator{\diag}{diag}

\newcommand{\ifbcdot}[1]{\ifblank{#1}{\cdot}{#1}}

\DeclarePairedDelimiterX\abs[1]{\lvert}{\rvert}{\ifbcdot{#1}}
\DeclarePairedDelimiterX\parens[1]{(}{)}{\ifbcdot{#1}}
\DeclarePairedDelimiterX\brk[1]{[}{]}{\ifbcdot{#1}}
\DeclarePairedDelimiterX\braces[1]{\{}{\}}{\ifbcdot{#1}}
\DeclarePairedDelimiterX\angles[1]{\langle}{\rangle}{\ifblank{#1}{\cdot,\cdot}{#1}}
\DeclarePairedDelimiterX\ip[2]{\langle}{\rangle}{\ifbcdot{#1},\ifbcdot{#2}}
\DeclarePairedDelimiterX\norm[1]{\lVert}{\rVert}{\ifbcdot{#1}}
\DeclarePairedDelimiterX\ceil[1]{\lceil}{\rceil}{\ifbcdot{#1}}
\DeclarePairedDelimiterX\floor[1]{\lfloor}{\rfloor}{\ifbcdot{#1}}

\DeclarePairedDelimiterXPP\trace[1]{\operatorname{Tr}}{(}{)}{}{\ifbcdot{#1}} 
\DeclarePairedDelimiterXPP\col[1]{\operatorname{col}}{\{}{\}}{}{\ifbcdot{#1}} 
\DeclarePairedDelimiterXPP\row[1]{\operatorname{row}}{\{}{\}}{}{\ifbcdot{#1}} 
\DeclarePairedDelimiterXPP\erf[1]{\operatorname{erf}}{(}{)}{}{\ifbcdot{#1}}
\DeclarePairedDelimiterXPP\erfc[1]{\operatorname{erfc}}{(}{)}{}{\ifbcdot{#1}}
\DeclarePairedDelimiterXPP\KLD[2]{D}{(}{)}{}{\ifbcdot{#1}\, \delimsize\|\, \ifbcdot{#2}} 
\DeclarePairedDelimiterXPP\op[2]{\operatorname{#1}}{(}{)}{}{#2} 

\newcommand{\convp}{\stackrel{\mathrm{p}}{\longrightarrow}}
\newcommand{\convas}{\stackrel{\mathrm{a.s.}}{\longrightarrow}}

\newcommand{\T}{^{\intercal}}
\newcommand{\setcomp}{^{\mathsf{c}}} 
\newcommand{\ud}{\,\mathrm{d}} 

\newcommand{\bzero}{\bm{0}}
\newcommand{\bone}{\bm{1}}

\DeclarePairedDelimiterXPP\indicate[1]{{\bf 1}}{\{}{\}}{}{\ifbcdot{#1}}

\newcommand{\indicatore}[1]{{\bf 1}_{#1}}
\newcommand{\ofrac}[1]{{\frac{1}{#1}}}

\newcommand{\ddfrac}[2]{{\dfrac{\mathrm{d} {#1}}{\mathrm{d} {#2}}}}
\newcommand{\ppfrac}[2]{\dfrac{\partial {#1}}{\partial {#2}}}

\providecommand\given{}

\DeclarePairedDelimiterX\Set[2]\{\}{%
\renewcommand\given{\SetSymbol[\delimsize]{#1}}
#2
}
\DeclarePairedDelimiterX\Setc[1]\{\}{%
\renewcommand\given{\SetSymbol{:}}
#1
}

\NewDocumentCommand\set{s o m}{%
	\IfBooleanTF#1%
	{\IfValueTF{#2}{\Set*{#2}{#3}}{\Setc*{#3}}}%
	{\IfValueTF{#2}{\Set{#2}{#3}}{\Setc{#3}}}%
}

\NewDocumentCommand{\evalat}{ s O{\big} m e{_^} }{%
\IfBooleanTF{#1}%
{\left. #3 \right|}{#3#2|}%
\IfValueT{#4}{_{#4}}%
\IfValueT{#5}{^{#5}}%
}

\providecommand\given{}
\DeclarePairedDelimiterXPP\cprob[1]{}(){}{
\renewcommand\given{\nonscript\,\delimsize\vert\allowbreak\nonscript\,\mathopen{}}%
#1%
}
\DeclarePairedDelimiterXPP\cexp[1]{}[]{}{
\renewcommand\given{\nonscript\,\delimsize\vert\allowbreak\nonscript\,\mathopen{}}%
#1%
}

\DeclareDocumentCommand \P { s e{_^} d() g } {%
	\mathbb{P}%
	\IfBooleanTF{#1}%
		{
			\IfValueT{#2}{_{#2}}%
			\IfValueT{#3}{^{#3}}%
			\IfValueTF{#5}{\cprob{#4 \given #5}}{\IfValueT{#4}{\cprob{#4}}}%
		}%
		{
			\IfValueT{#2}{_{#2}}%
			\IfValueT{#3}{^{#3}}%
			\IfValueTF{#5}{\cprob*{#4 \given #5}}{\IfValueT{#4}{\cprob*{#4}}}%
		}%
}

\DeclareDocumentCommand \E { s e{_^} o g } {%
	\mathbb{E}%
	\IfBooleanTF{#1}%
		{
			\IfValueT{#2}{_{#2}}%
			\IfValueT{#3}{^{#3}}%
			\IfValueTF{#5}{\cexp{#4 \given #5}}{\IfValueT{#4}{\cexp{#4}}}%
		}%
		{
			\IfValueT{#2}{_{#2}}%
			\IfValueT{#3}{^{#3}}%
			\IfValueTF{#5}{\cexp*{#4 \given #5}}{\IfValueT{#4}{\cexp*{#4}}}%
		}%
}

\ExplSyntaxOn
\NewDocumentCommand \dist {m o o} {%
\mathrm{#1}\left(%
	\IfValueT{#3}{%
		\tl_if_blank:nTF{ #3 }{\cdot\, \middle|\, }{#3\, \middle|\, }%
	}
	\IfValueT{#2}{#2}%
\right)%
}
\ExplSyntaxOff

\newcommand{\N}[2]{\dist{\calN}[#1,\, #2]}

\NewDocumentCommand {\cbrace} {t+ D[]{black} D(){\widthof{#5}} m m } {%
	\begingroup%
		\color{#2}
		\IfBooleanTF{#1}{%
			\overbrace{#4}^%
		}{
			\underbrace{#4}_%
		}%
		{\parbox[c]{#3}{\centering\footnotesize{#5}}}%
	\endgroup%
}

\let\oldforall\forall
\renewcommand{\forall}{\oldforall \, }

\let\oldexist\exists
\renewcommand{\exists}{\oldexist \, }

\graphicspath{{./Figures/}{./figures/}}
\DeclareDocumentCommand{\includeCroppedPdf}{ o O{./Figures/} m }{
	\IfFileExists{#2#3-crop.pdf}{}{%
		\immediate\write18{pdfcrop #2#3.pdf #2#3-crop.pdf}}%
	\includegraphics[#1]{#2#3-crop.pdf}
}

\makeatletter
\newcommand*{\addFileDependency}[1]{
  \typeout{(#1)}
  \@addtofilelist{#1}
  \IfFileExists{#1}{}{\typeout{No file #1.}}
}
\makeatother

\definecolor{gray90}{gray}{0.9}

\ifx\nohighlights\undefined
	\newcommand{\red}[1]{{\color{red} #1}}
	
	\newcommand{\msout}[1]{\text{\color{green} \sout{\ensuremath{#1}}}}
	\newcommand{\del}[1]{{\color{green}\ifmmode \msout{#1}\else\sout{#1}\fi}}
\else
	\newcommand{\red}[1]{#1}
	
	\newcommand{\msout}[1]{#1}
	\newcommand{\del}[1]{#1}
\fi

\newcommand{\hhide}[1]{}

\ifx\diagnoselabel\undefined
	\relax
\else
	\makeatletter
	\def\@testdef #1#2#3{%
		\def\reserved@a{#3}\expandafter \ifx \csname #1@#2\endcsname
			\reserved@a  \else
			\typeout{^^Jlabel #2 changed:^^J%
				\meaning\reserved@a^^J%
				\expandafter\meaning\csname #1@#2\endcsname^^J}%
			\@tempswatrue \fi}
	\makeatother
\fi

\newacronym{KLD}{KL divergence}{Kullback–Leibler divergence}
\newacronym{pdf}{pdf}{probability density function}

\renewcommand{\exp}{e}
\newcommand{\eps}{\epsilon}
\newcommand{\TV}[2]{\mathrm{TV}({#1}, {#2})}
\newcommand{\ChiSq}[2]{\chi^2({#1}\, \|\, {#2})}
\newcommand{\nChiSq}[2]{\hat{\chi}^2_m({#1}\, \|\, {#2})}
\newcommand{\Fdiv}[3]{{D_{#1}({#2}\, \|\, {#3})}}
\renewcommand{\KLD}[2]{D_{\mathrm{KL}}\left(\ifbcdot{#1}\, \|\, \ifbcdot{#2}\right)} 

\begin{document}

\title{On the Relationship Between Information-Theoretic Privacy Metrics And Probabilistic Information Privacy}
%
\author{Chong~Xiao~Wang and Wee~Peng~Tay,~\IEEEmembership{Senior Member,~IEEE} 
	\thanks{The authors are with the School of Electrical and Electronic Engineering, Nanyang Technological University, Singapore. E-mails: \texttt{\{wangcx, wptay\}@ntu.edu.sg}}
}



\maketitle


\begin{abstract}
Information-theoretic (IT) measures based on $f$-divergences have recently gained interest as a measure of privacy leakage as they allow for trading off privacy against utility using only a single-value characterization. However, their operational interpretations in the privacy context are unclear. In this paper, we relate the notion of probabilistic information privacy (IP) to several IT privacy metrics based on $f$-divergences. We interpret probabilistic IP under both the detection and estimation frameworks and link it to differential privacy, thus allowing a precise operational interpretation of these IT privacy metrics. We show that the $\chi^2$-divergence privacy metric is stronger than those based on total variation distance and Kullback-Leibler divergence. Therefore, we further develop a data-driven empirical risk framework based on the $\chi^2$-divergence privacy metric and realized using deep neural networks. This framework is agnostic to the adversarial attack model. Empirical experiments demonstrate the efficacy of our approach.
\end{abstract}

\begin{IEEEkeywords}
	Inference privacy, privacy measure, $f$-divergence, differential privacy, $\chi^2$-divergence.
\end{IEEEkeywords}

\section{Introduction} \label{sect:intro}

The past decades have witnessed the proliferation of digital services such as cloud computing, which necessitates the collection of prodigious amounts of data from a myriad of sources. The concomitant risk of exposing sensitive information arouses the antipathy of data owners towards external access to their data. For example, studies have shown that users' personal information such as sexual orientation and political affiliation can be accurately inferred from their activities on social networking platforms \cite{FirGolElo:J14}. Data providers must privatize or sanitize the data to mitigate the tension between the need to share data and the need to protect sensitive information \cite{AgrSri:J00}. 

Data privacy involves the proper collection and dissemination of data in ways that conceal the identity or attribute of any individual datum while inference privacy \cite{CalFaw:C12, SunTay:J19a, SunTay:J19b, SunTayHe:J18, HeTayHua:J19, WanSonTay:J21} seeks to prevent the disclosure of sensitive information that is statistically dependent on the original data. The key distinction is that inference privacy is completely built upon a statistical inference framework, while ingredients of data privacy can be partially or totally non-stochastic. For both cases, a major challenge in developing privacy-preserving methodologies is to formally quantify the amount of privacy leakage, given all possible auxiliary information the adversary may have. This quantification plays a crucial role in designing privatization schemes as an indicator of the necessary amount of perturbation needed for a desirable level of privacy protection.

\subsection{Privacy Metrics}
Privacy notions that have gained wide visibility trace back to the concept of group-based anonymization, which hides individual records by reducing the granularity of data in a database. A popular technique is $k$-anonymity \cite{Swe:J02}, which guarantees that the identity of an individual whose data is contained in a database is indistinguishable from at least $k-1$ other individual participants when projected on the quasi-identifiers. However, attackers can still make inferences about sensitive values that exhibit homogeneity within an anonymized group. Subsequently, $\ell$-diversity \cite{MacGehKif:C06} is proposed to overcome the weakness of the anonymized database by additionally requiring the sensitive fields in an equivalence class to have at least $\ell$ well-represented values to maintain diversity. One problem with $\ell$-diversity is that it does not consider semantic meanings of sensitive values and hence is not immune to attacks with global knowledge about the sensitive attributes. The definition of $t$-closeness \cite{RebForFer:J10} refines $\ell$-diversity by taking into account the distributions of the sensitive attributes in an equivalence class and the whole database. 

Over the past decade, differential privacy (DP) \cite{DwoMcsNis:C06,Dwo:C08,DwoLei:C09,DwoRot:J14} has emerged out of attempts to withhold individual information when releasing aggregate information about a database. Owing to its rigorous approach and formal privacy guarantees, DP has become the mainstream data privacy metric. It formalizes the idea that the presence or absence of an individual in a database does not appreciably affect the distribution of a randomized inquiry. Compared to $k$-anonymity and $\ell$-diversity, which are semantic, DP is algorithmic and provides semantic privacy guarantees \cite{LiQarSu:J11,EkeOngLiu:J22}. One of the extraordinary characteristics of DP is that it abstracts away the attacker's auxiliary information about the data, and DP is thus proof against an attacker with arbitrary side information \cite{KaiSmi:J14}. However, enforcing this strict guarantee comes with a price. A differentially private algorithm in practice can significantly distort data, thus diminishing the overall utility of the privatized results \cite{WanLeiFie:J16, SorDomSan:J17, SunTay:J19b}.

It should be noted that DP is independent of the data distribution. Going beyond this, many privacy works leverage the distribution of the data to obtain interesting results. For instance, references \cite{SorDom:C13,DomSor:J15} relate $t$-closeness to DP by making assumptions about the prior and posterior views of the data. The work \cite{DimBelZha:J17} demonstrates that under proper choices of the prior, responding to queries using samples from the posterior is sufficient to guarantee DP, and the work \cite{LiQarSu:C13} generalizes DP by choosing prior distribution families.

Because the data distribution is often available to the attacker as side information, privacy mechanisms can take advantage of the uncertainty of the data in a probabilistic manner. For example, Bayesian DP proposed by \cite{TriFal:C20} calibrates noise perturbation to the data distribution to provide practical DP guarantees. Quantifiers from information theory \cite{CovTho:B91} that measure the uncertainty of a random variable from observing another random variable become a natural choice to formalize the measure of privacy leakage as well as utility. The reader is referred to the survey \cite{BloGunYen:J21} for a detailed history of the field. Works like \cite{CalFaw:C12, SanRajPoo:J13, SanRajMoh:J13, MakSalFaw:C14} cast the privacy-utility trade-off as a modified rate-distortion problem \cite{CovTho:B05} or the opposite of the information bottleneck problem \cite{NatPerWil:C19}, in which finding the privatization scheme is formulated as an optimization over a privacy-assuring probabilistic mapping. The most well-known information-theoretic (IT) privacy metrics include mutual information, total variation distance \cite{RasGun:J20}, chi-square information and maximal correlation \cite{MakFaw:C13, AsoDiaLin:J16, CalMak:J17, WanCal:C17, WanVoCal:J19}, which are the subjects of our study.

There is a growing interest in IT privacy metrics as each typically uses a single-value characterization of privacy leakage (e.g., mutual information), whereas the number of constraints to formulate DP is contingent on the size of data, thus making it unwieldy in optimization frameworks. Due to their concise formulations, IT privacy metrics can be combined with a utility measure as a loss function for finding an optimal sanitizer while maintaining computational tractability. Therefore, IT privacy metrics are more accessible to many application domains that emphasize optimal privacy-utility trade-off. On the other hand, DP suffers from several practical problems and limitations \cite{CliTas:J13}. For example, employing DP as a privacy measure for learning an arbitrary sanitizer \cite{KaiOhVis:J16} requires the data distribution to be known. The differentially private mechanism of adding Laplacian noise can significantly decrease the utility. In contrast, in practical cases where the data is continuous and high-dimensional and its distribution is unavailable, it is possible to derive an estimate of an IT privacy metric from a finite number of samples.

On the downside \cite{Issa:J20}, IT privacy metrics do not come with a cogent operational interpretation. Although operational interpretations of some IT privacy metrics like mutual information do arise in transmission and compression settings and are related to statistical dependency between variables, they are not explicit operational interpretations like those provided by privacy notions like DP and information privacy (IP) \cite{CalFaw:C12, SunTay:J19a, SunTay:J19b}. This paper aims to bridge this gap.

\subsection{Contributions}
The goal of this paper is to provide an interpretation of IT privacy metrics formed by $f$-divergences. This is achieved by relating to the notion of probabilistic IP \cite{SunTayHe:J18}, which confines an adversary's posterior belief about the private variable with high probability. While it has been shown that DP can bound IT privacy metrics (e.g., $\eps$-DP ensures $\eps$-mutual information privacy) \cite{AlvAndCha:C11}, how IT privacy metrics can imply (weak) DP has not been identified yet. The authors in \cite{SarSan:C14, WanYinZha:J16} investigated the relationship between mutual information and DP based on their impact on data distortion. To the best of our knowledge, our work is the first paper that examines the connections between $f$-divergence IT privacy metrics and probabilistic IP (cf. \cref{cor:strong_itp2pip}) and thus weak DP (cf. \cref{lem:IP-DP}).
Our contributions are summarized as follows:
\begin{itemize}
	\item We review the probabilistic IP concept, which is consistent with an axiomatic view of a leakage measure. We show that probabilistic IP implies weak DP. Probabilistic IP is premised on a Bayesian model, which allows us to exploit the adversary's uncertainty about data. The key to probabilistic IP is restricting the coverage of privacy protection to typical scenarios (which contain the events that are likely to happen). We show how probabilistic IP is related to the decision error under the detection framework and the mean square estimation error under the estimation framework.
	
	\item We derive the relationship of several IT privacy metrics formed by $f$-divergences to probabilistic IP. The $f$-divergences we study are the total variation (TV) distance, Kullback-Leibler (KL) divergence and $\chi^2$-divergence. We show that the IT privacy metric that is strongest amongst them is the $\chi^2$-divergence privacy metric. 
	
	\item We consider practical cases where data distribution is not available and propose a statistically consistent estimator of the $\chi^2$-divergence. Based on that, we develop a data-driven framework for learning a neural network sanitizer, which can be instantiated appropriately depending on the problem domain. 
\end{itemize}
The focus of this paper is on the interpretation of IT privacy metrics via their relationships to probabilistic IP. It is expected that some of our results are useful in studying privacy-utility trade-offs. The latter study is interesting future work and beyond the scope of the current paper.

The rest of the paper is organized as follows. In \cref{sect:probabilistic IP}, we bring in the notion of probabilistic IP and derive its properties. In \cref{sect:itp2pip}, we characterize IT privacy metrics using probabilistic IP. In \cref{sect:dd_privacy}, we present an estimate of the $\chi^2$-divergence which converges in the large sample size regime and propose a data-driven privacy-preserving framework using the $\chi^2$-divergence privacy metric. In \cref{sect:expt}, we conduct experiments for privacy-utility trade-off. Finally, we make conclusions in \cref{sect:conclusion}.

\emph{Notations:}
We use capital letters like $X$ to denote random variables or vectors, and lowercase letters like $x$ for deterministic scalars or vectors. Throughout this paper, all random variables are defined on the same probability space with probability measure $\P$. We use $\E[X]\coloneqq\int X\ud{\P}$ to denote the expectation of $X$ and $\E[X \mid Y]$ is the conditional expectation. We assume that every random variable has a (generalized) probability density function (pdf) (for discrete random variables, this specializes to a probability mass function). We use $p_X(\cdot)$ to denote the pdf of $X$, and $p_{X \mid Y}(\cdot \mid \cdot)$ to denote the conditional pdf of $X$ given $Y$. We use $X\sim p$ to say that the random variable $X$ follows a pdf $p$. We use $\E_{X\sim p}[X]\coloneqq\int x p(x)\ud{x}$ to emphasize that the expectation is \gls{wrt} $X$ with pdf $p$. We use $\circ$ to denote function composition. The Cartesian product of two sets $\calA$ and $\calB$ are denoted as $\calA\times\calB$. The indicator function $\indicatore{\calA}(x)$ takes value $1$ if and only if $x$ belongs to set $\calA$. $\Gamma\setcomp$ denotes the complement of the set $\Gamma$. We denote $\abs{a}$ as the absolute value of $a$. The inverse function of a function $f$ is $f^{-1}$. The logarithm $\log$ is the natural logarithm.

\section{Probabilistic Information Privacy}\label{sect:probabilistic IP}

In this section, we review the probabilistic IP definition and concept \cite{CalFaw:C12, SunTayHe:J18}. We characterize the properties of probabilistic IP under a statistical framework and show that probabilistic IP implies DP with high probability. Our goal is to relate probabilistic IP with IT privacy metrics that are based on $f$-divergences. These are introduced in \cref{sect:itp2pip}.

Consider a probability space $(\Omega,\calF,\P)$, where $\Omega$ is the sample space, $\calF$ is a $\sigma$-algebra of events and $\P$ is a probability measure. A random element $X$ is a measurable function from $(\Omega,\calF)$ to $(\calX,\calB(\calX))$, where $\calX$ is a topological space $X$ takes values in and $\calB(\calX)$ denotes the Borel $\sigma$-algebra generated by the open sets of $\calX$. Recall that for any subset $A\subset\Omega$, $X(A)=\set{X(\omega)\in\calX \given \omega\in A}$ is the image of $A$ under $X$. For any subset $B\subset\calX$, the inverse set map $X^{-1}(B)=\set{\omega\in\Omega\given X(\omega)\in B}$.

We use a random element $S$ taking values in some set $\calS$ to typify the private variable to be protected. A random element $X\in\calX$ denotes the raw data, which is supposed to be released but is correlated with $S$. Releasing $X$ will inevitably disclose information about $S$. To preserve the privacy of $S$, we let $X$ pass through a noisy channel $p_{Y \mid X}$. This generates a sanitized variable $Y\in\calY$ to replace $X$ as the released data. The process of generating $Y$ from $X$ is called the privatization mechanism or data sanitization. Note that $S$, $X$ and $Y$ form a Markov chain $S-X-Y$.

When sanitizing $X$ to produce $Y$, the utility of $Y$ should also be taken into consideration. However, measuring utility is not the focus of this paper, and we simply quantify it by the empirical risk in \cref{sect:expt}. The discussion of privacy definitions involves the random elements $S$ and $Y$ only. 

In this paper, for simplicity, we assume that all random elements have probability density functions or probability mass functions (i.e., there exists a dominating probability measure \gls{wrt} we can take Radon-Nikodym derivatives).
Accordingly, $p_S(\cdot)$ and $p_Y(\cdot)$ denote the marginal distributions of $S$ and $Y$, respectively.
We assume that $p_S(s)>0$ for all $s\in\calS$.


\subsection{Definition of probabilistic IP}

A privacy metric provides a formal measure of the amount of privacy ``leakage'' when publishing the sanitized variable. In a general statistical framework, the prior distribution (before the release of any information) of the private variable $S$ is known to an adversary, which constitutes the adversary's side information. 
For each $(s,y)\in (\calS,\calY)$, the \emph{relative disparity} between the posterior belief (after observing $Y=y$) and the prior about $S=s$ is defined as
\begin{align*}
	d(s,y)=\frac{p_{S\mid Y}(s\mid y)}{p_S(s)}.
\end{align*}
For $\eps>0$, $S$ given $Y$ achieves $\eps$-information privacy ($\eps$-IP) \cite{CalFaw:C12, SunTayHe:J18} if for almost surely all $(s,y)$, we have
\begin{align}\label{eq:ip}
	\exp^{-\eps}\leq d(s,y) \leq\exp^{\eps},
\end{align}
where $\eps$ is called the privacy budget. The privacy budget limits the adversary's posterior belief about $S$ when observing $Y$. We note that $\eps$-IP provides the worst-case privacy guarantee in at least two senses. First, inequality \cref{eq:ip} must hold for every $s\in\calS$, meaning that the privacy for almost surely every $s\in\calS$ is protected. Second, inequality \cref{eq:ip} requires that the bounds hold for almost surely every possible sanitization outcome $y\in\calY$, even if $y$ occurs only with very low probability. This can be unwieldy in many practical learning settings. For example, the privatization mechanism designer may not have global knowledge about the population of $S$ or $Y$ but has access to only data samples. Moreover, an excessive utility trade-off may be needed to account for the rare cases of $(s,y)$.

Probabilistic IP is a relaxation of $\eps$-IP by imposing the privacy constraint \cref{eq:ip} on the most probable occurrences (which are referred to as typical scenarios). As a consequence, it is possible but unlikely for an adversary to gain information about the private variable $S$ from observing the sanitized variable $Y$. We give the formal definition of probabilistic IP, or, equivalently, $(\eps,\delta)$-IP as follows.

\begin{Definition}[$(\eps,\delta)$-IP; \cite{SunTayHe:J18}]
	\label{def:probabilistic IP}
	For $\eps>0$ and $0\leq\delta\leq1$, we say $S$ given $Y$ achieves $(\eps,\delta)$-IP if
	\begin{align}\label{eq:probabilistic IP}
		\P( \set*{\omega\in\Omega \given \exp^{-\eps}\leq d(S(\omega),Y(\omega)) \leq\exp^{\eps}} )\geq 1-\delta,
	\end{align}
	and achieves \emph{strong} $(\eps,\delta)$-IP if 
	\begin{align}\label{eq:strong_PIP}
		\P( \bigcap_{s\in\calS} \set*{\omega\in\Omega \given \exp^{-\eps}\leq d(s,Y(\omega)) \leq\exp^{\eps}} )\geq 1-\delta.
	\end{align}
\end{Definition}
There is a subtle but non-trivial difference between \cref{eq:probabilistic IP,eq:strong_PIP} in \cref{def:probabilistic IP}. The event in \cref{eq:probabilistic IP} includes the randomness of both $S$ and $Y$, whereas, in \cref{eq:strong_PIP}, the event of interest is \gls{wrt} the randomness of $Y$ only (i.e., the former is a union of events while the latter is an intersection of events). The motivation behind \cref{eq:strong_PIP} is the observation that in a majority of practical problems 
we desire that a sanitized variable $Y$ does not disclose information about $S$, regardless of the realization of $S$. In this case, we only require that this happens with high probability. 

By taking $\delta\to 0$, $(\eps,\delta)$-IP degenerates to $\eps$-IP. Either decreasing $\eps$ or $\delta$ yields a stronger privacy guarantee.

To facilitate our analysis, we define two useful ``tail'' events in which the sanitized variable $Y$ leaks information about $S$:
\begin{align}
	&\calL_{\eps}
	=\set*{\omega\in\Omega\given d\left(S(\omega),Y(\omega)\right)<\exp^{-\eps}}, \label{Leps}\\
	&\calR_{\eps}
	=\set*{\omega\in\Omega\given d\left(S(\omega),Y(\omega)\right)>\exp^{\eps}}. \label{Reps}
\end{align}
Note that $(\eps,\delta)$-IP is equivalent to $\P(\calL_{\eps}\cup\calR_{\eps})\leq\delta$. Since $Y(\calL_{\eps} \cup \calR_{\eps})$ is the set of values $Y(\omega)$ in $\calY$ for some $\omega\in \calL_{\eps} \cup \calR_{\eps}$, we have 
\begin{align}\label{SIPC}
Y^{-1}\circ Y(\calL_{\eps}\cup\calR_{\eps}) &= 
\bigcup_{s\in\calS} \set*{\omega\in\Omega \given \exp^{-\eps}\leq d(s,Y(\omega)) \leq\exp^{\eps}}\setcomp, 
\end{align} 
and strong $(\eps,\delta)$-IP in \cref{eq:strong_PIP} is equivalent to 
\begin{align*}
	\P(Y^{-1}\circ Y(\calL_{\eps}\cup\calR_{\eps}))\leq\delta.
\end{align*}
It is obvious that strong $(\eps,\delta)$-IP implies $(\eps,\delta)$-IP because
\begin{align*}
	\P(Y^{-1}\circ{Y}(\calL_{\eps}\cup\calR_{\eps}))
	\geq\P(\calL_{\eps}\cup\calR_{\eps}).
\end{align*}
If $S$ given $Y$ achieves $(\eps,\delta)$-IP, it also achieves $(\eps',\delta')$-IP for any $\eps'>\eps$ and $\delta'>\delta$ because $\calL_{\eps'}$ (resp. $\calR_{\eps'}$) is a subset of $\calL_{\eps}$ (resp. $\calR_{\eps}$) for $\eps'\geq\eps$.

We wish to make connections between probabilistic IP and weak DP or $(\eps,\delta)$-DP since DP is deemed a gold standard within the privacy research community. We recall the concept of $(\eps,\delta)$-DP, whose goal is to simultaneously withhold information about an individual record in a database when releasing aggregate information about the database. A randomized query is differentially private if it is almost equally likely to be from any two databases that differ in a single individual data record. In the following, we adopt a stronger notion of neighbors in our inference framework. 

\begin{Definition}[$(\eps,\delta)$-Differential privacy]\label{def:DP}
	Any $s\in\calS$ and $s'\in\calS$ are said to be neighbors if they take distinct values.
	We say $S$ given $Y$ achieves $(\eps,\delta)$-DP if for every pair of neighbors $s,s'\in\calS$ and all $A\in\calB(\calY)$, we have
	\begin{align*}
		\P(Y\in A\given S=s)
		\leq\exp^{\eps}\P(Y\in A\given S=s')+\delta.
	\end{align*}
	If $\delta=0$, we say that $S$ given $Y$ achieves $\eps$-DP.
\end{Definition}

\begin{Remark}
If $\calS \subset \Real^n$ (e.g., in a database), the typical definition of neighbors $s=(s_1,s_2,\dots,s_n)$ and $s'=(s'_1,s'_2,\dots,s'_n)$ in the DP framework require that $s$ and $s'$ differ only in one component, i.e., $s_i\ne s'_i$ for some $i$ and $s_j=s'_j$ for all $j\ne i$. In our inference framework, $\calS$ is not necessarily embedded in an $n$-dimensional vector space. Hence, we consider any distinct $s$ and $s'$ to be neighbors. Nevertheless, our framework can also accommodate database privacy using the usual definition of neighbors in DP.
\end{Remark}

For every run of the privatization algorithm $p_{Y \mid X}$, $\eps$-DP ensures that $Y$ is almost equally likely to be observed \emph{on every pair} of neighboring private data, simultaneously. In practice, $\eps$-DP can be too strong to satisfy in some scenarios. A commonly used relaxation is to allow a small error probability such that it is possible but unlikely that ex post facto an observation of $Y$ will be much more or much less likely to be generated when $S=s$ than when $S=s'$ (cf. \cite[Lemma 3.17]{DwoRot:J14}). 

As opposed to DP, which is independent of the prior distribution of $S$, probabilistic IP makes use of $p_S$ to model the side information of an adversary \cite{DimBelZha:J17, TriFal:C20}. In addition, the interpretation of $\delta$ in DP is somewhat problematic due to taking the probability space over the privatization mechanism. As pointed out by \cite{LiQarSu:J11}, the probability that a privacy breach occurs is not bounded by $\delta$ in DP. In contrast, $\delta$ in probabilistic IP explicitly amounts to the probability over the ``tail'' scenarios out of the coverage of privacy protection.

It is easy to see that $\eps$-IP immediately leads to $2\eps$-DP \cite{SunTay:J19b}. In what follows, we show that strong $(\eps,\delta)$-IP can guarantee a certain level of $(\eps,\delta)$-DP.
\begin{Lemma}\label{lem:IP-DP}
	Suppose $\alpha=\inf_{s\in\calS}p_S(s)>0$. If $S$ given $Y$ achieves strong $(\eps,\delta)$-IP, it is also $(2\eps,\delta/\alpha)$-DP. 
\end{Lemma}
\begin{IEEEproof}
	Let $\Psi=Y^{-1}\circ Y(\calL_{\eps}\cup\calR_{\eps})$. For $y\in Y(\Psi\setcomp)$ and any neighbors $s,s'\in\calS$ with $s\neq s'$, we have
	\begin{align*}
		\frac{p_{Y \mid S}(y \mid s)}{p_{Y \mid S}(y \mid s')}
		=\frac{p_{S \mid Y}(s \mid y)}{p_S(s)}\frac{p_S(s')}{p_{S \mid Y}(s' \mid y)}
		\leq\exp^{2\eps}.
	\end{align*}
	Therefore, for any $B\subset\Psi\setcomp$, we have
	\begin{align}
		\P(B\given S=s)\leq\exp^{2\eps}\P(B\given S=s').\label{Bexps}
	\end{align}
	On the other hand, we have
	\begin{align}
		\P(\Psi\given S=s)
		\leq\frac{\P(\Psi\cap S^{-1}(s))}{\P(S=s)}
		\leq\frac{\P(\Psi)}{\P(S=s)}\leq\delta/\alpha.\label{Psidelta}
	\end{align}
    Finally, for any $A\in\calB(\calY)$, we have
    \begin{align*}
    	\P(Y^{-1}(A)\given S=s)
    	&=\P(Y^{-1}(A)\cap\Psi\setcomp\given S=s)+\P(Y^{-1}(A)\cap\Psi\given S=s)\nn
    	&\leq\exp^{2\eps}\P(Y^{-1}(A)\cap\Psi\setcomp\given S=s')+\P(\Psi\given S=s)\nn
    	&\leq\exp^{2\eps}\P(Y^{-1}(A)\given S=s')+\P(\Psi\given S=s),
    \end{align*}
	where the last equality follows from \cref{Bexps}. From \cref{Psidelta}, the proof is complete.
\end{IEEEproof}

From the proof of \cref{lem:IP-DP}, we also have that $(\eps,\delta)$-IP ensures $2\eps$-DP with probability $1-\delta$ (\gls{wrt} the randomness over $S$ and $Y$).

\subsection{Error Bounds}

The goal of invoking a privacy definition is to limit an adversary's capability of inferring $S$ based on $Y$. Therefore, a quantitative characterization of this capability is important to justify the appropriateness of the privacy definition. We show that $(\eps,\delta)$-IP indeed lower-bounds the detection error and estimation error of $S$. 
The following \cref{lemma:detec_err} provides a non-trivial bound to the probability of error under the detection framework when enforcing $(\eps,\delta)$-IP. 

\begin{Lemma}\label{lemma:detec_err}
	Suppose $\calS$ and $\calY$ are finite alphabets, and $S$ given $Y$ achieves $(\eps,\delta)$-IP.
	Then, for any decision rule $\gamma:\calY\to\calS$, we have  
	\begin{align*}
		\P(\gamma(Y)\neq S)
		\geq 1-\delta-\exp^{\eps}\max_{s\in\calS}p_S(s).
	\end{align*}
\end{Lemma}
\begin{IEEEproof}
	It is known that the maximum a posteriori rule minimizes $\P(\gamma(Y)\neq S)$, i.e., the optimal decision rule $\gamma$ is given by
	\begin{align*}
		\gamma(y)=\argmax_{s\in\calS} p_{S \mid Y}(s \mid y),\ \forall y\in\calY.
	\end{align*}
	Let $\Gamma_y=S(Y^{-1}(y)\cap\calR_{\eps})$ for $y\in\calY$. Firstly, we have
	\begin{align}\label{max_ineq_a}
		&\sum_{y\in\calY}p_Y(y)\max_{s\in\Gamma_y} p_{S \mid Y}(s \mid y)\\
		&\leq\sum_{y\in\calY}\sum_{s\in\Gamma_y} p_{S,Y}(s,y)\nn
		&=\sum_{y\in\calY}\sum_{s\in\Gamma_y}\P(S^{-1}(s)\cap Y^{-1}(y))
		=\P(\calR_{\eps})\leq\delta, \nonumber
	\end{align}
	where the last inequality is due to $\P(\calR_{\eps})\leq\P(\calL_{\eps}\cup\calR_{\eps})\leq\delta$.
	Secondly, we have
	\begin{align}\label{max_ineq_b}
		&\sum_{y\in\calY}p_Y(y)\max_{s\in\Gamma_y\setcomp}p_{S \mid Y}(s \mid y)\\
		&\leq\sum_{y\in\calY}p_Y(y)\max_{s\in\Gamma_y\setcomp}\set*{\exp^{\eps}p_S(s)}
		\leq\exp^{\eps}\max_{s\in\calS}p_S(s).\nonumber
	\end{align}
	Finally, the proof is completed by noting that
	\begin{align*}
		\sup_{\gamma}\P(\gamma(Y)=S)
		=\sum_{y\in\calY}p_Y(y)\max_{s\in\calS}p_{S \mid Y}(s \mid y)
		\leq\cref{max_ineq_a}+\cref{max_ineq_b}.
	\end{align*}
\end{IEEEproof}
Either decreasing $\eps$ or $\delta$ elevates the lower bound of the error probability, which suggests a lower accuracy for the Bayes classifier. This observation is consistent with the claim that a smaller $\eps$ or $\delta$ provides stronger privacy protection. In the extreme case where $\eps=\delta=0$, it is no surprise that the bound reaches the largest Bayes error of $1-\max_{s\in\calS}p_S(s)$.
%

Next, we provide a bound for the estimation error when enforcing $(\eps,\delta)$-IP for continuous $S$ and $Y$. Note that estimation error is defined \gls{wrt} the variable range while $(\eps,\delta)$-IP is not. To relate them, we need to assume a regularity condition.

\begin{Lemma}\label{lemma:est_err}
	Suppose $\calS \subset \bbR_{\geq0}$ and $\calY\subset\bbR$.
	Let $\calM=(\calL_{\eps}\cup\calR_{\eps})\setcomp$ and $\Gamma_y=S(Y^{-1}(y)\cap\calM)$.
	Suppose for $y\in Y(\calM)$ and $\alpha\in\{1,2\}$, the following regularity condition holds:
	\begin{align}\label{rgl_cond}
		\E[S^{\alpha}]
		=\E[S^{\alpha}\given S^{-1}(\Gamma_y)].
	\end{align} 
	If $S$ given $Y$ achieves $(\eps,\delta)$-IP, then for any estimator $\gamma:\calY\to\calS$, we have
	\begin{align*}
		\E[(S-\gamma(Y))^2]
		\geq(1-\delta){\exp^{-2\eps}}\E[S^2]-{\exp^{2\eps}}\E[S]^2.
	\end{align*}
\end{Lemma}
\begin{IEEEproof}
Firstly, we have
\begin{align}
	\E[\E[S\indicatore{\calM}\given Y]^2]
	&=\int_{Y(\calM)}\left(\int_{\Gamma_y}sp_{S \mid Y}(s \mid y)\ud{s}\right)^2p_Y(y)\ud{y}\nn
	&\leq\int_{Y(\calM)}\left(\exp^{\eps}\int_{\Gamma_y}sp_S(s)\ud{s}\right)^2p_Y(y)\ud{y}\nn
	&\leq\exp^{2\eps}\int_{Y(\calM)} \E[S\given\Gamma_y]^2p_Y(y)\ud{y}\nn
	&=\exp^{2\eps}\E[S]^2\int_{Y(\calM)}p_Y(y)\ud{y}\nn
	&\leq\exp^{2\eps}\E[S]^2.\label{total_var_a}
\end{align}
Secondly, we have
\begin{align}
	\E[(S\indicatore{\calM})^2]
	&=\int_{\calM}s^2p_{S,Y}(s,y)\ud{s}\ud{y}\nn
	&=\int_{Y(\calM)}p_Y(y)\int_{\Gamma_y}s^2p_{S \mid Y}(s \mid y)\ud{s}\ud{y}\nn
	&\geq\exp^{-\eps}\int_{Y(\calM)}p_Y(y)\int_{\Gamma_y}s^2p_S(s)\ud{s}\ud{y}\nn
	&=\exp^{-\eps}\E[S^2]\int_{Y(\calM)}\int_{\Gamma_y}p_Y(y)p_S(s)\ud{s}\ud{y}\nn
	&\geq\exp^{-2\eps}\E[S^2]\int_{\calM}p_{S,Y}(s,y)\ud{s}\ud{y}\nn
	&\geq(1-\delta)\exp^{-2\eps}\E[S^2].\label{total_var_b}
\end{align}
Finally, we have
\begin{align}
	\E[(S-\gamma(Y))^2]
	&\geq\E[\left(S\indicatore{\calM}-\gamma(Y)\indicatore{\calM}\right)^2]\nn
	&\geq\E[\left(S\indicatore{\calM}-\E[S\indicatore{\calM}\given Y]\right)^2]\nn
	&=\E[(S\indicatore{\calM})^2]-\E[\E[S\indicatore{\calM}\given Y]^2].\label{total_var_c}
\end{align}
The proof is completed by substituting \cref{total_var_a,total_var_b} into \cref{total_var_c}.
\end{IEEEproof}

To interpret the regularity condition in \cref{lemma:est_err}, note for $y\in Y(\calM)$,
\begin{align*}
	\Gamma_y=\set*{s\given\exp^{-\eps}\leq d(s,y) \leq \exp^{\eps}}\subset\calS.
\end{align*}
Thus, $\Gamma_y$ contains all points in $\calS$ that are protected by $(\eps,\delta)$-IP when conditioned on $Y=y$. The regularity condition \cref{rgl_cond} ensures that the first and second moments of $S$ on $\Gamma_y$ are consistent with that over $\Gamma_y\setcomp$. The regularity condition is always satisfied for strong $(\eps,\delta)$-IP because $\Gamma_y=\calS$ for $y\in Y(\calM)$ by \cref{def:probabilistic IP} and hence $S^{-1}(\Gamma_y)=\Omega$. When $S$ is independent of $Y$, the estimation error bound reaches its maximum value (which equals the variance of $S$). One can enlarge this error bound by decreasing $\delta$ or $\eps$ to provide stronger privacy protection.

\section{From IT Privacy Metrics to probabilistic IP}
\label{sect:itp2pip}

In this section, we present the relationship of several well-known IT privacy metrics with probabilistic IP, to provide insights into the operational principles of IT privacy metrics as privacy measures.

We begin by reviewing the definitions of the IT privacy metrics studied in this paper. First, we introduce $f$-divergences \cite{Csis:J64,AliSil:J66}, which are a general class of statistical distances measuring the divergence between two probability distributions over the same probability space.

\begin{Definition}[$f$-divergence]
	Let $\bbP$ and $\bbQ$ be two probability measures over a sample space $\Omega$ such that $\bbP$ is absolutely continuous \gls{wrt} $\bbQ$. For a convex function $f:[0,\infty)\to\bbR$ such that $f(1)=0$, the $f$-divergence from the reference measure $\bbQ$ to $\bbP$ is
	\begin{align}\label{Fdiv}
		\Fdiv{f}{\bbP}{\bbQ}
		=\int_{\Omega}f\left(\ddfrac{\bbP}{\bbQ}\right)\ud{\bbQ}.
	\end{align}
\end{Definition}
Many of the common statistical divergences are special cases generated by different choices of function $f$. For example, total variation (TV) distance, Kullback-Leibler (KL) divergence and $\chi^2$-divergence are associated with generating functions $f(x)=\abs{x-1}$, $f(x)=x\log x$ and $f(x)=x^2-1$, respectively. Given two density functions $p$ and $q$ over $\calZ$, the total variation distance between $p$ and $q$ is
\begin{align*}
	\TV{p}{q}=\int_{\calZ}\abs{p(z)-q(z)}\ud{z},
\end{align*}
the KL divergence between $p$ and $q$ is
\begin{align*}
	\KLD{p}{q}=\int_{\calZ}p(z)\log\frac{p(z)}{q(z)}\ud{z},
\end{align*}
and the $\chi^2$-divergence between $p$ and $q$ is
\begin{align*}
	\ChiSq{p}{q}=\int_{\calZ}\frac{p(z)}{q(z)}p(z)\ud{z}-1.
\end{align*}
We restrict our discussion to the above three types of $f$-divergences. IT privacy metrics formed by the $f$-divergences between the joint distribution and the product of the marginal distributions of the private variable and the sanitized variable are widely used to quantify inference privacy \cite{CalFaw:C12,WanVoCal:J19,WanCal:C17,RasGun:J20,ZhaGor:J22}.

\begin{Definition}[$f$-divergence privacy metrics]
Denote 
\begin{align*}
	q_{S,Y}(s,y)=p_S(s)p_Y(y),\
	\forall (s,y)\in\calS\times\calY.
\end{align*}
For $\eta\geq 0$, we say that 
\begin{itemize}
	\item $S$ given $Y$ satisfies $\eta$ $f$-divergence privacy if
	\begin{align}\label{eq:itp}
		\Fdiv{f}{p_{S,Y}}{q_{S,Y}}
		=\E[\Fdiv{f}{p_{Y\mid S}(\cdot\mid S)}{p_Y(\cdot)}]\leq\eta.
	\end{align}
	\item $S$ given $Y$ satisfies strong $\eta$ $f$-divergence privacy if for almost surely all $s\in\calS$,
	\begin{align}\label{eq:strong_itp}
		\Fdiv{f}{p_Y(\cdot)}{p_{Y \mid S}(\cdot \mid s)}\leq\eta.
	\end{align}
\end{itemize}
\end{Definition}
Strong $f$-divergence privacy is tailored for privacy problems with $|\calS|<\infty$ because only in this case is \cref{eq:strong_itp} numerically tractable for \emph{every} $s\in\calS$.

Note that \cref{eq:itp} with the KL divergence is the mutual information between $S$ and $Y$, which is a quantity of statistical dependence between $S$ and $Y$ \cite{CovTho:B91}. The reference distribution $q_{S,Y}$ for the $f$-divergences in \cref{eq:itp} is chosen according to this analogy. Conversely, the choice of the reference distribution in \cref{eq:strong_itp} does not follow this rule. As shown in \cref{cor:strong_itp2pip}, this choice leads to the conclusion that strong $f$-divergence privacy implies strong $(\eps,\delta)$-IP. 
In what follows, we present the main result of this paper: $f$-divergence privacy implies $(\eps,\delta)$-IP.

\begin{Theorem}\label{thm:itp2pip}
	The following $\eta$ $f$-divergence privacies based on TV distance, KL divergence and $\chi^2$-divergence, imply $(\eps,\delta)$-IP, for any $\eps > 0$ and $\delta$ specified by $\eta$ and $\eps$ as follows.
	\begin{enumerate}[(a)]
		\item\label[claim]{it:TV} If $\TV{p_{S,Y}}{q_{S,Y}}\leq\eta$, then $S$ given $Y$ achieves $(\eps,\delta)$-IP with $\delta=\dfrac{\eta}{1-\exp^{-\eps}}$.
		\item\label[claim]{it:KL} If $\KLD{p_{S,Y}}{q_{S,Y}}\leq\eta$, then $S$ given $Y$ achieves $(\eps,\delta)$-IP with $\delta=\zeta(\eps)+\zeta(-\eps)$, where
		\begin{align*}
			\zeta(\eps)=\sup\set*{p\in\left[0,1\right]\given (1-p)\log\frac{1-p}{\exp^{\eps}-p}\leq\eta-\eps}.
		\end{align*}
		\item\label[claim]{it:Chi} If $\ChiSq{p_{S,Y}}{q_{S,Y}}\leq\eta$, then $S$ given $Y$ achieves $(\eps,\delta)$-IP with
		\begin{align*}
			\delta=\dfrac{\exp^{-\eps}\eta}{(\exp^{-\eps}-1)^2+\eta}
			+\dfrac{\exp^{\eps}\eta}{(\exp^{\eps}-1)^2+\eta}.
		\end{align*}
	\end{enumerate}
\end{Theorem}
\begin{IEEEproof}
The proofs of \cref{it:TV,it:KL,it:Chi} are presented in \cref{proof:itp2pip_a,proof:itp2pip_b,proof:itp2pip_c}, respectively.
\end{IEEEproof}

\Cref{thm:itp2pip} gives a characterization of $f$-divergence privacy from the perspective of probabilistic IP, thus allowing us to assign the operational interpretations of probabilistic IP to these $f$-divergence privacies. For a given level of $f$-divergence privacy, \cref{thm:itp2pip} casts light on which level $\eps$-IP or $\eps$-DP is protected with high probability. Note the $\delta$ in $(\eps,\delta)$-IP resulting from $f$-divergence privacy is coupled with $\eps$. For a fixed $\eta$, increasing $\eps$ decreases $\delta$, and for a fixed $\eps$, increasing $\eta$ increases $\delta$. Although $\eps$ can be evaluated at any positive value, the resulting $\delta$ may become trivial if $\delta\geq 1$.

Taking the results in \cref{thm:itp2pip} further, we show that strong $f$-divergence privacy implies strong $(\eps,\delta)$-IP.
\begin{Corollary}\label{cor:strong_itp2pip}
	Suppose $\abs{\calS}<\infty$ and $\Fdiv{f}{p_Y}{p_{Y \mid S}(\cdot \mid s)}\leq\eta$ for all $s\in\calS$. For $\eps>0$, $S$ given $Y$ achieves strong $\left(\eps,\delta\abs{\calS}\right)$-IP, with the same $\delta$ given in \cref{thm:itp2pip} for total variation distance, KL divergence and $\chi^2$-divergence, respectively.
\end{Corollary}
\begin{IEEEproof}
	For each $s\in\calS$, let
	\begin{align*}
		\Gamma_s
		=\set*{\omega\in\Omega\given d(s,Y(\omega))\geq\exp^{\eps}}
		\cup\set*{\omega\in\Omega\given d(s,Y(\omega))\leq\exp^{-\eps}}.
	\end{align*}
	Retracing the proof steps of \cref{thm:itp2pip}, it can be deduced that if $\Fdiv{f}{p_Y}{p_{Y \mid S=s}}\leq\eta$ for each of the $f$-divergences in \cref{thm:itp2pip}, we have $\P(\Gamma_s)\leq\delta$ with $\delta$ given in \cref{thm:itp2pip}. Note this is true only if $p_{Y \mid S=s}$ acts as the reference distribution. Recall that $S$ given $Y$ achieves strong $(\eps,\delta)$-IP if 
	\begin{align*}
		\P(Y^{-1}\circ Y(\calL_{\eps}\cup\calR_{\eps}))\leq\delta.
	\end{align*}
	
	For any $\omega\in Y^{-1}\circ Y(\calL_{\eps}\cup\calR_{\eps})$, there exists $s\in\calS$ such that $d(s,Y(\omega))\geq\exp^{\eps}$ or  $d(s,Y(\omega))\leq\exp^{-\eps}$. Therefore, we must have
	\begin{align}
		Y^{-1}\circ Y(\calL_{\eps}\cup\calR_{\eps})\subset\bigcup_{s\in\calS}\Gamma_s. \label{strongIPsubset}
	\end{align}
	As a result, we have
	\begin{align*}
		\P(Y^{-1}\circ Y(\calL_{\eps}\cup\calR_{\eps}))
		\leq\sum_{s\in\calS}\P(\Gamma_s)
		\leq\delta\abs{\calS}.
	\end{align*}
	The proof is now complete.
\end{IEEEproof}

\begin{Remark}
	Following \cref{thm:itp2pip}, one may be interested in whether it is possible to translate probabilistic IP into $f$-divergence privacy. The answer is positive for the total variation distance as shown in \cref{lemma:tv2pip} below. However, the question remains to be explored for the other $f$-divergences.
	\begin{Lemma}\label{lemma:tv2pip}
		If $S$ given $Y$ achieves $(\eps,\delta)$-IP, we have
		\begin{align*}
			\TV{p_{S,Y}}{q_{S,Y}}
			\leq 2(\exp^{\eps}-1+\delta).
		\end{align*}
	\end{Lemma}
	\begin{IEEEproof}
		See \cref{proof:tv2pip}.
	\end{IEEEproof}
\end{Remark}

Apart from the IT privacy metrics based on $f$-divergences, maximal correlation \cite{Gel:J1941,Hir:J35} defined in \cref{def:max_corr} below has also been extensively employed as a measure of privacy leakage from an estimation-theoretic point of view \cite{Calmon:C13,AsoAla:C15,AsoDiaLin:J16,LiGamal:J18}.
\begin{Definition}[The Hirschfeld-Gebel\'ein-Renyi Maximal Correlation]
\label{def:max_corr}
Let $Z\in\calZ$ and $W\in\calW$ be jointly distributed random variables. Denote $\calH(p_Z)=\set*{f \given\E_{Z\sim p_Z}[f(Z)]=0,\E_{Z\sim p_Z}[f(Z)^2]=1}$.
The maximal correlation between $Z$ and $W$ is
\begin{align*}
	\rho_m(Z,W)
	=\sup_{\substack{f\in\calH(p_Z) \\ g\in\calH(p_W)}}\E[f(Z)g(W)].
\end{align*}	
\end{Definition}

The following result shows the relationship between $\chi^2$-divergence and maximal correlation.
\begin{Lemma}
	\label{lemma:max_corr}
	The following inequalities hold:
	\begin{align*}
		\frac{\ChiSq{p_{S,Y}}{q_{S,Y}}}{\min\{\abs{\calS},\abs{\calY}\}-1}
		\leq\rho_m(S,Y)^2\leq\ChiSq{p_{S,Y}}{q_{S,Y}}.
	\end{align*}
\end{Lemma}
\begin{IEEEproof}
If both $\calS$ and $\calY$ are infinite alphabets, the lower bound holds vacuously. Therefore, we assume at least one is finite. The rest of the proof is in \cref{proof:lemma:max_corr}.
\end{IEEEproof}
The IT privacy metrics and maximal correlation are formal measures of the statistical dependence between $S$ and $Y$. They possess desirable properties such as vanishing if and only if $S$ and $Y$ are independent (perfect privacy). The usage of IT privacy metrics in a privacy configuration is typically to form a loss function along with a utility measure for optimizing a privatization mechanism.

With the availability of several privacy metrics studied in this paper, a natural question arises: which privacy metric should one choose? While there does not exist a unified answer as the choice often depends on the problem domain, it is possible to compare these privacy metrics in a universal sense as follows \cite{SunTay:J19b}.

\begin{Definition}\label{defn:strong_p}
	We say type A privacy metric is stronger than type B privacy metric if for \textbf{any} valid privacy budget $\eta$, there exists $\eta'$ such that any $S$ given $Y$ that achieves $\eta'$ type A privacy also satisfies $\eta$ type B privacy. If two privacy metrics are stronger than each other, we say they are equivalently strong.
\end{Definition}

From the Pinkster's inequality \cite{CovTho:B91}, we have 
\begin{align*}
	\TV{p_{S,Y}}{q_{S,Y}}^2\leq\KLD{p_{S,Y}}{q_{S,Y}}.
\end{align*}
From Jensen's inequality, we have
\begin{align*}
	\KLD{p_{S,Y}}{q_{S,Y}}
	=\E[\log\left(\frac{p_{S,Y}(S,Y)}{q_{S,Y}(S,Y)}\right)]
	&\leq\log\E[\frac{p_{S,Y}(S,Y)}{q_{S,Y}(S,Y)}]	
	=\log(\ChiSq{p_{S,Y}}{q_{S,Y}}+1).
\end{align*}
Using \cref{defn:strong_p}, $\chi^2$-divergence privacy metric is, therefore, stronger than the privacy metrics formed by KL divergence and total variation distance. Furthermore, \cref{lemma:max_corr} indicates that $\chi^2$-divergence and maximal correlation are equivalently strong if the private variable is discrete. In general, $\chi^2$-divergence is the strongest privacy metric amongst the IT privacy metrics referenced in this section. 

\subsection{Translating to Weak DP}
In \cref{cor:strong_itp2pip}, it has been shown that strong IT privacy metrics imply strong probabilistic IP, and \cref{lem:IP-DP} shows that strong probabilistic IP implies weak DP (when the private variable $S$ has finite support). By chaining these two results, we immediately obtain a lower bound of weak DP that is guaranteed by the IT privacy metric. In what follows, we illustrate this lower bound using an example of the Gaussian mechanism of DP \cite{DwoRot:J14}.

\begin{figure}[!htb]
	\centering
	\includegraphics[scale=0.5]{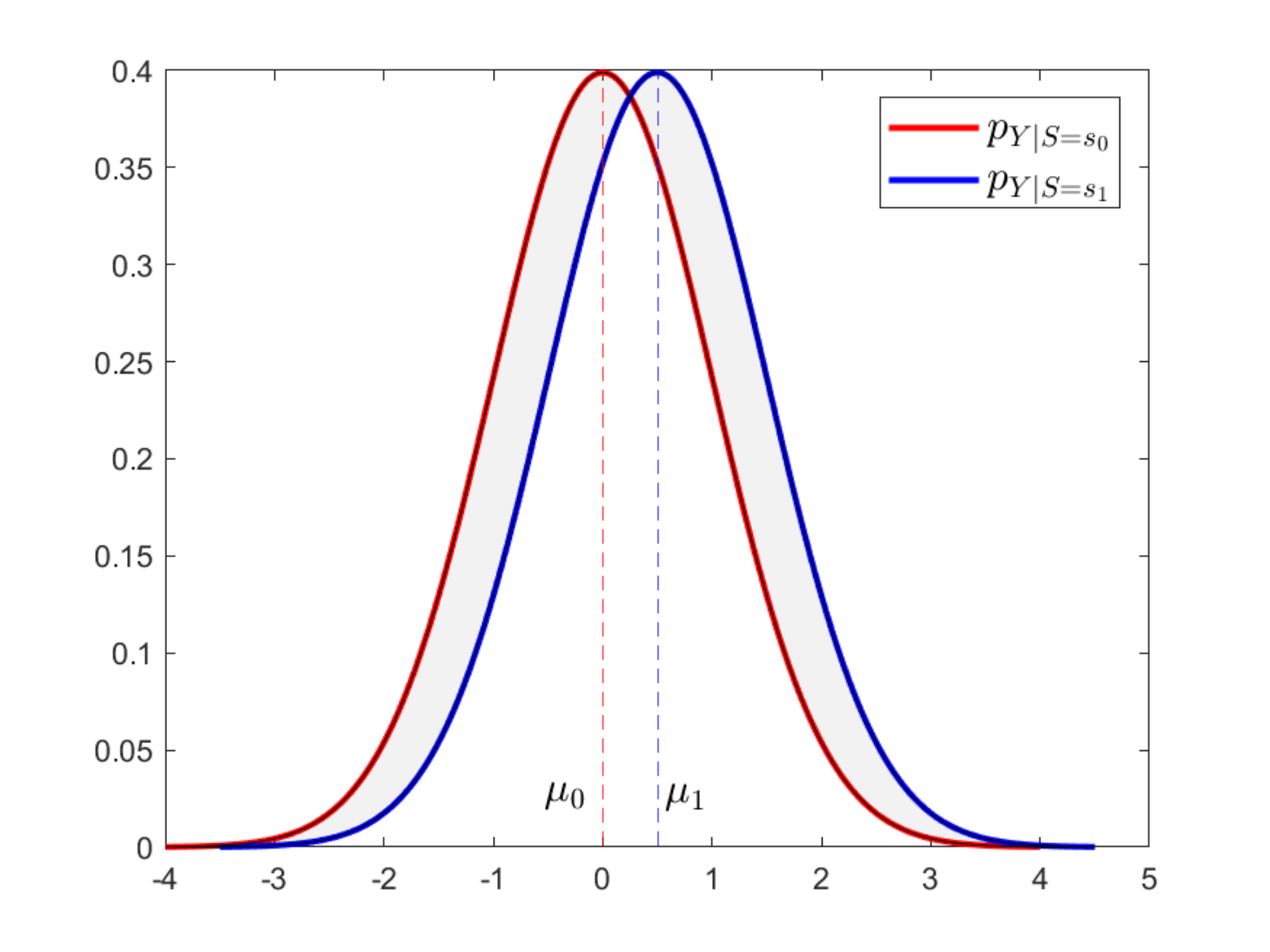}
	\caption{Gaussian mechanism. When $\mu_0$ and $\mu_1$ are close, it is almost equally likely for most realizations of $Y$ (except for the tail part) to be generated from $S=s_0$ and $S=s_1$.}
	\label{fig:gaussian_mechanism}
\end{figure}

Consider a private variable $S=\set{s_0,s_1}$ and a continuous sanitized variable $Y$ whose distribution is specified by
\begin{align*}
	p_{Y \mid S=s_0}=\N{\mu_0}{\sigma^2},\nn
	p_{Y \mid S=s_1}=\N{\mu_1}{\sigma^2}.
\end{align*}
From the Gaussian mechanism, $S$ given $Y$ achieves $(\eps,\delta)$-DP if
\begin{align*}
	\sigma^2=\frac{2(\mu_0-\mu_1)^2}{\epsilon^2}\log(1.25/\delta).
\end{align*}
For an illustration, see \cref{fig:gaussian_mechanism}.

Now we fix $\eps$ and $\delta$ for the Gaussian mechanism, and compute the $\chi^2$-divergence privacy for $S$ and $Y$. 
Note that DP disregards the prior distribution of $S$. The $\chi^2$-divergence between two normal distributions can be computed analytically:
\begin{align*}
	\ChiSq{p_{Y \mid S=s_0}}{p_{Y \mid S=s_1}}
	=\mathrm{exp}\left(\frac{(\mu_0-\mu_1)^2}{\sigma^2}\right).
\end{align*}
From $p_Y=p_S(s_0)p_{Y \mid S=s_0}+p_S(s_1)p_{Y \mid S=s_1}$, it can be verified that
\begin{align*}
	&\ChiSq{p_Y}{p_{Y \mid S=s_0}}
	=p_S(s_1)^2\ChiSq{p_{Y \mid S=s_1}}{p_{Y \mid S=s_0}},\nn
	&\ChiSq{p_{Y}}{p_{Y \mid S=s_1}}
	=p_S(s_0)^2\ChiSq{p_{Y \mid S=s_0}}{p_{Y \mid S=s_1}}.
\end{align*}

\begin{figure}[!htb]
	\centering
	\begin{subfigure}{.49\textwidth}
		\centering
		\includegraphics[width=\textwidth]{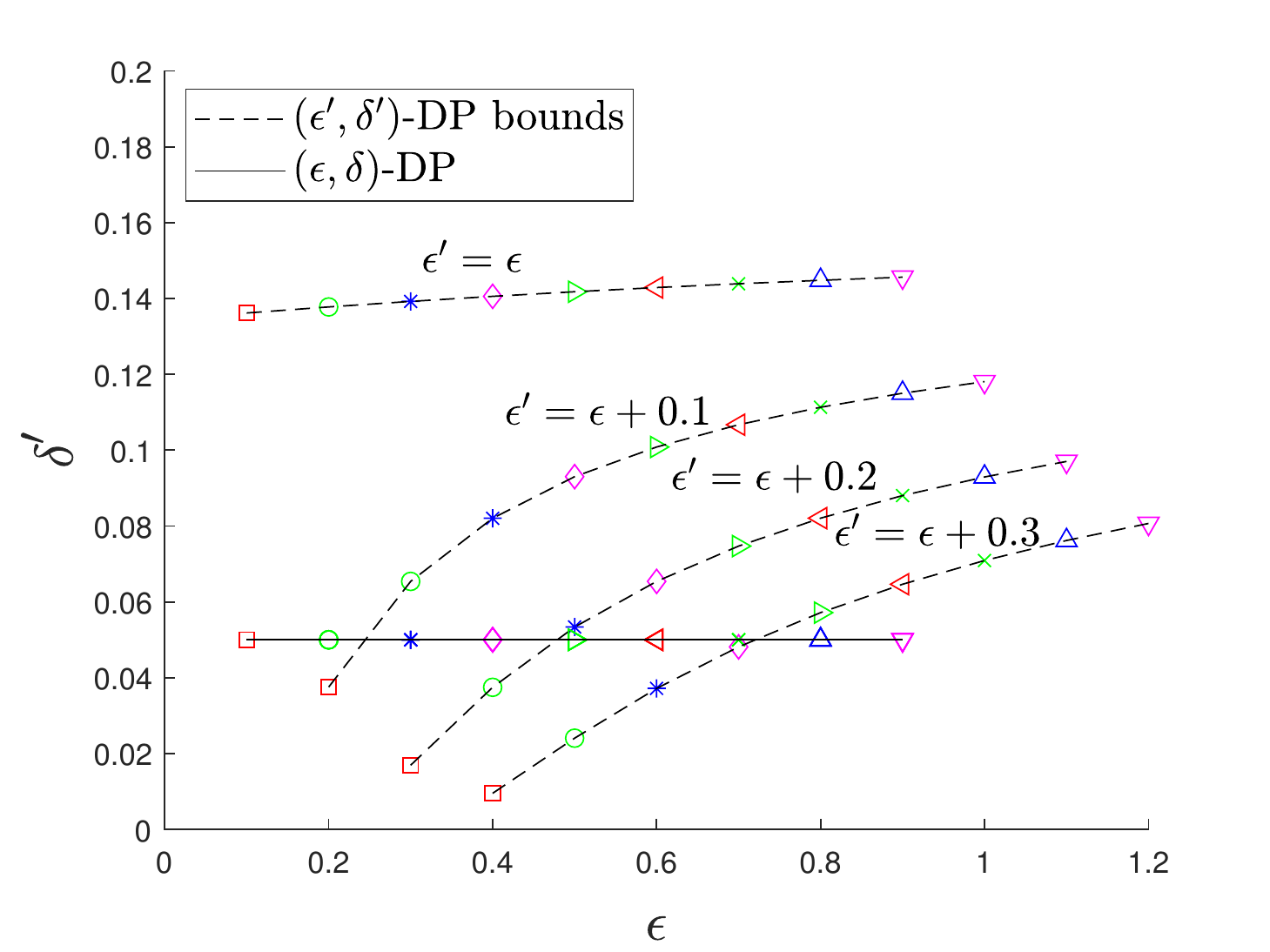}
		\caption{$(\eps',\delta')$-DP bounds with varying $\eps$ and $\delta=0.05$.}
		\label{fig:dp2pip2dp_a}
	\end{subfigure}
	\begin{subfigure}{.49\textwidth}
		\centering
		\includegraphics[width=\textwidth]{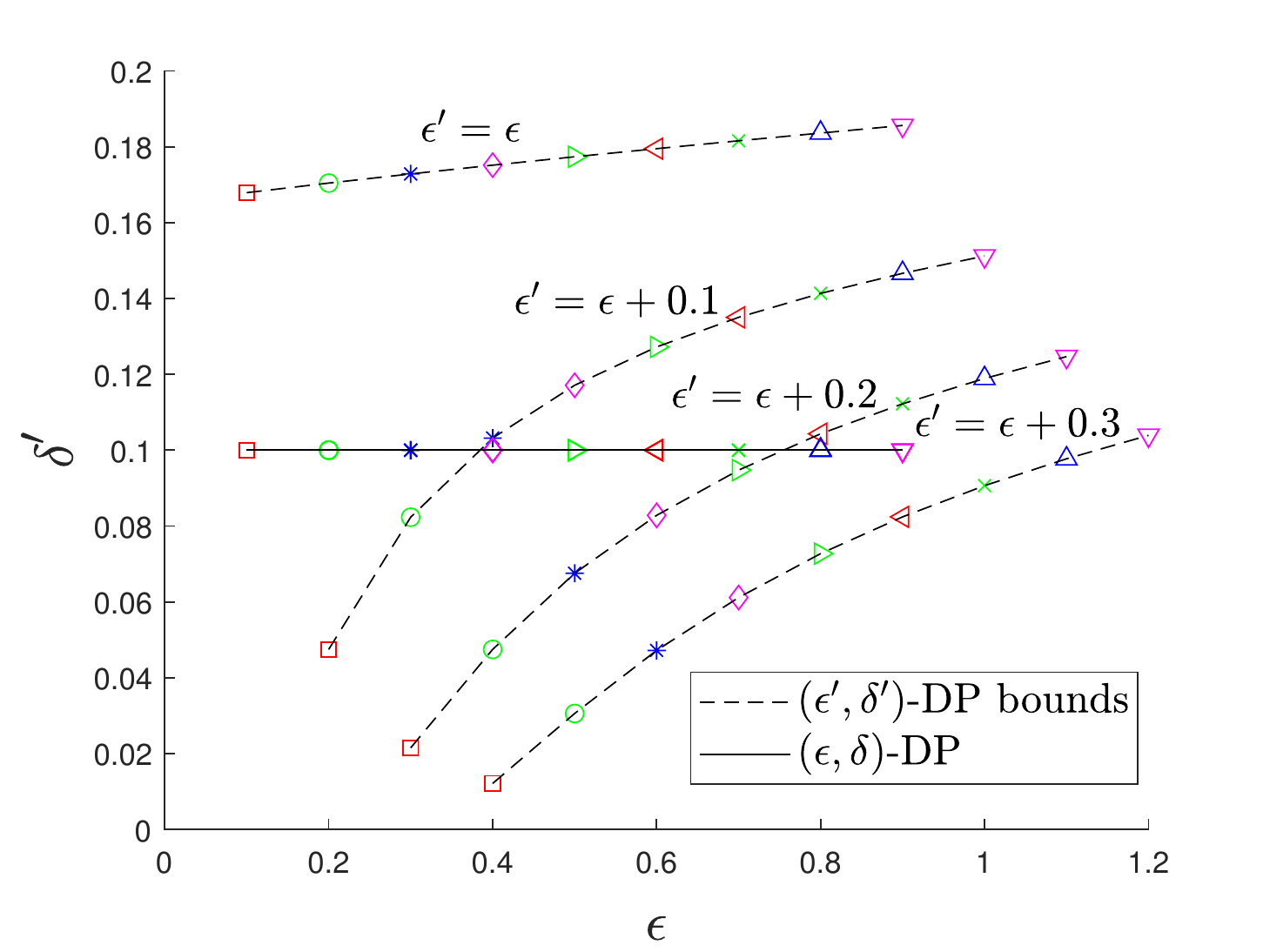}
		\caption{$(\eps',\delta')$-DP bounds with varying $\eps$ and $\delta=0.1$.}
		\label{fig:dp2pip2dp_b}
	\end{subfigure}
	\begin{subfigure}{.49\textwidth}
		\centering
		\includegraphics[width=\textwidth]{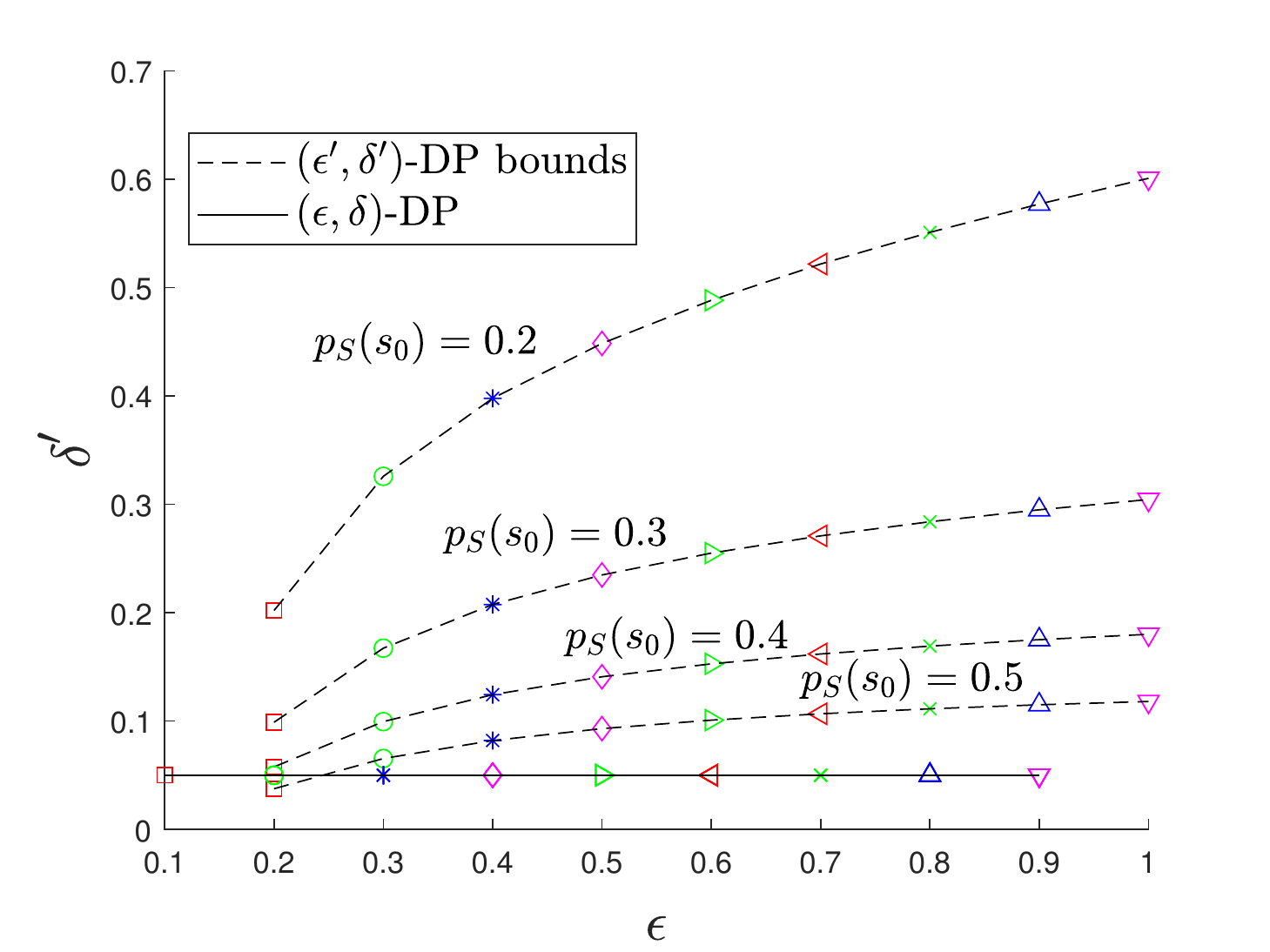}
		\caption{$(\eps',\delta')$-DP bounds with varying prior $p_S$.}
		\label{fig:dp2pip2dp_c}
	\end{subfigure}
	\caption{$(\eps',\delta')$-DP bounds derived from the $\chi^2$-divergence privacy for $(\eps,\delta)$-DP Gaussian mechanism.}
\end{figure}

Based on the $\chi^2$-divergence privacy determined by $(\eps,\delta)$-DP, we firstly use \cref{cor:strong_itp2pip} to quantify the strong IP, and then apply \cref{lem:IP-DP} to compute the $(\eps',\delta')$-DP bound. We compare the derived $(\eps',\delta')$-DP bounds with the baseline $(\eps,\delta)$-DP. In \cref{fig:dp2pip2dp_a,fig:dp2pip2dp_b}, we set $\delta=0.1$ and $\delta=0.05$, respectively, and vary $\epsilon$ from $0.1$ to $1.2$, while fixing $p_S(s_0)=p_S(s_1)=0.5$. Note \cref{thm:itp2pip} indicates that $\delta'$ is a function of $\eps'$ for the $(\eps',\delta')$-DP bound, and we can evaluate $\eps'$ at any value. Letting $\eps'$ be the sum of $\eps$ and a small positive value, we obtain $\delta'$. It can be seen that $\delta'$ decreases as $\eps'$ increases, implying that weaker privacy protection always comes with a higher probability. The $(\eps',\delta')$ bound becomes tighter when $(\eps,\delta)$ is closer to $(0,0)$. In \cref{fig:dp2pip2dp_c}, we vary $p_S$ to verify its impact on DP. The results are consistent with \cref{lem:IP-DP}, which states the level of DP under probabilistic IP is related to $\min_{s\in\calS}p_S(s)$. The bound tends to be looser when the prior of $S$ is unbalanced.

\section{Data-driven Privacy Metric}
\label{sect:dd_privacy}

In this section, we propose a practical implementation of $\chi^2$-divergence based on a variational form and show that the proposed empirical estimate is asymptotically consistent. This lays a foundation of the data-driven privacy-preserving framework in \cref{sect:expt}.

One prominent advantage of an $f$-divergence privacy metric is that it can be estimated from data without the need to estimate the data distribution, which is particularly useful for high-dimensional and continuous data. This stands in striking contrast to DP, which is unmanageable in such cases. The variational form views $f$-divergence from an optimization perspective, for which approximation is feasible by restricting the search function space to be from a parametric family represented by neural networks.

In what follows, we review the dual representation of $\chi^2$-divergence and propose a tighter and regularized representation. Let $p$ and $q$ be two probability distributions over $\calZ$. 
A common variational formulation of \cref{Fdiv} is obtained via the Legendre-Fenchel duality \cite{SreGol:J22}. The conjugate of the convex function $f:[0,\infty)\to\bbR$ in \cref{Fdiv} is defined as
\begin{align*}
	f^*(w)=\sup_{z\in\bbR_+}\set{zw-f(z)}.
\end{align*}
Note $f^{**}=f$ when $f$ is convex and closed. This yields a dual representation of $f$-divergence \cite{SreGol:J22}:
\begin{align*}
	\Fdiv{f}{p}{q}
	=\sup_{g\in\calH}\set*{\E_{Z\sim p}[g(Z)]-\E_{Z\sim q}[f^*(g(Z))]},
\end{align*}
where $\calH$ includes all measurable functions from $\calZ$ to $\bbR$ such that the last expectation term is finite. In particular, the $\chi^2$-divergence admits the following the dual representation \cite{SreGol:J22}:
\begin{align}\label{eq:dual_chisq}
	\ChiSq{p}{q}
	=\sup_{g\in\calH}\set*{\E_{Z\sim p}[g(Z)]-\E_{Z\sim q}[g(Z)+g(Z)^2/4]},
\end{align}
where the supremum is achieved at $g(z)=2\left(\dfrac{p(z)}{q(z)}-1\right)$. 

In \cref{prop:var_chisq}, we present an improved variational form of $\chi^2$-divergence \cite{PolyansklyWu2022}. We note that the optimal $g$ in \cref{eq:dual_chisq} must satisfy the regularization $\E_{Z\sim p}[g(Z)]=1$, whereas this is not required in \cref{eq:var_chisq}. 
\begin{Proposition}\label{prop:var_chisq}
	Let $\calH=\set*[\vert]{g:\calZ\to\bbR\given 0<\E_{Z\sim q}[g(Z)^2]<\infty}$. Assume $q>0$ almost surely. $\chi^2$-divergence admits the following variational form:
	\begin{align}\label{eq:var_chisq}
		\ChiSq{p}{q}
		=\sup_{g\in\calH}\frac{\left(\E_{Z\sim p}[g(Z)]-\E_{Z\sim q}[g(Z)]\right)^2}{\E_{Z\sim q}[g(Z)^2]}.
	\end{align}
\end{Proposition}
\begin{IEEEproof}
	From the Cauchy-Schwarz inequality, we have
	\begin{align*}
		\E_{Z\sim q}[\left(1-\frac{p(Z)}{q(Z)}\right)g(Z)]^2
		&\leq\E_{Z\sim q}[\left(1-\frac{p(Z)}{q(Z)}\right)^2]\E_{Z\sim q}[g(Z)^2]\nn
		&=\ChiSq{p}{q}\E_{Z\sim q}[g(Z)^2],
	\end{align*}
	where the inequality becomes equality when $g(z)\propto 1-\dfrac{p(z)}{q(z)}$ for $z\in\calZ$ almost everywhere.
\end{IEEEproof}

Now suppose we are given two sets of samples $\{w_i\}_{i=1}^m$ and $\{z_i\}_{i=1}^m$ drawn independently from $p$ and $q$, respectively, and we want to estimate the $\chi^2$-divergence \cref{eq:var_chisq} using these samples. To ensure computational tractability, we let $\calH=\set*{g_{\bphi}\given\phi\in\Phi}$ in which $g_{\bphi}$ is a neural network function parameterized by trainable weights vector $\bphi\in\Phi$. Replacing the expectations in \cref{eq:var_chisq} with their respective sample averages, $\ChiSq{p}{q}$ can be estimated as
\begin{align}\label{eq:chisq_est}
	\nChiSq{p}{q}
	=\sup_{\bphi\in\Phi}\frac{\left(\ofrac{m}\sum_{i=1}^mg_{\bphi}(w_i)-\ofrac{m}\sum_{i=1}^mg_{\bphi}(z_i)\right)^2}
	{\ofrac{m}\sum_{i=1}^mg_{\bphi}(z_i)^2+\lambda_m},
\end{align}
where $\lambda_m\to 0$ is a regularization term for countering a vanishing denominator. 

The convergence of the empirical estimates to their corresponding population statistics with increasing sample size is important to justify the method. We show that the estimate \cref{eq:chisq_est} converges to \cref{eq:var_chisq} in probability (denoted as ``$\convp$'') if some mild assumptions are satisfied.

\begin{Theorem}\label{thm:chisq_est_covp}
	The estimate $\nChiSq{p}{q}\convp\ChiSq{p}{q}$ as $m\to\infty$ if the following conditions hold:
	\begin{enumerate}[(a)]
		\item\label{it:g} There exists $\bphi\in\Phi$ such that $g_{\bphi}(z)\propto 1-\dfrac{p(z)}{q(z)}$.
		\item\label{it:gs} $g_{\bphi}(z)$ is smooth \gls{wrt} $\bphi\in\Phi$ and continuous \gls{wrt} $z\in\calZ$.
		\item\label{it:gz} $g_{\bphi}(z)\neq 0$ for almost everywhere $z\in\calZ$.
		\item\label{it:compact} $\Phi$ and $\calY$ are compact.
	\end{enumerate}
\end{Theorem}
\begin{IEEEproof}
	From condition~\ref{it:g}, in \cref{eq:var_chisq}, we can restrict to $g = g_{\bphi}$ for some $\bphi\in\Phi$. Let its objective function be denoted as $\gamma(\bphi)$ and let $\gamma_m(\bphi)$ be the objective function of \cref{eq:chisq_est}. It suffices to prove
	\begin{align}\label{eq:generic_convp}
		\sup_{\bphi\in\Phi}\gamma_m(\bphi)\convp\sup_{\bphi\in\Phi}\gamma(\bphi).
	\end{align}
	From the generic uniform convergence theorem \cite[Theorem 1]{New:J91}, \cref{eq:generic_convp} is ensured by the following conditions:
	\begin{enumerate}[i.]
		\item $\Phi$ is compact.
		\item $\gamma_m(\bphi)\convas\gamma(\bphi)$ for all $\bphi\in\Phi$.
		\item $\gamma_m(\bphi)$ is stochastically equicontinuous for all $m\geq 1$, i.e., for any $\eps>0$, there exists $\sigma>0$ such that
		\begin{align}\label{eq:stoch_equicont}
			\lim_{m\to\infty}\P(\sup_{\norm{\bphi-\bphi'}<\delta}\abs*{\gamma_m(\bphi)-\gamma_m(\bphi')}>\eta)
		\end{align}
	\end{enumerate}
	Note condition i is given by condition \ref{it:compact} and condition ii follows from the strong law of large numbers. We only need to prove condition iii, which needs an auxiliary \cref{lemma:bounded_seq}.
	\begin{Lemma}
		\label{lemma:bounded_seq}
		If conditions \ref{it:gs}, \ref{it:gz} and \ref{it:compact} are satisfied, there exists a sequence of random variables $(B_m)_{m\geq1}$ and a constant $b<\infty$ such that 
		\begin{align*}
			&\lim_{m\to\infty}\P(B_m-b>\eps)=0,
			\intertext{for any $\eps>0$ and}
			&\abs{\lambda_m(\bphi)-\lambda_m(\bphi')}
			\leq B_m\norm{\bphi-\bphi'},
		\end{align*}
		for all $\bphi,\bphi'\in\Phi$, in which $\norm{}$ is the Euclidean norm.
	\end{Lemma}
	\begin{IEEEproof}
		See \cref{proof:bounded_seq}.
	\end{IEEEproof}
	Applying \cref{lemma:bounded_seq} to \cref{eq:stoch_equicont} and letting $\delta=\dfrac{\eta}{b+\eps}$ with $\eps>0$, we have
	\begin{align*}
		\cref{eq:stoch_equicont}&\leq\lim_{m\to\infty}\P(\sup_{\norm{\bphi-\bphi'}<\delta}B_m\norm*{\bphi-\bphi'}>\eta)\nn
		&\leq\lim_{m\to\infty}\P(B_m\delta>\eta)\nn
		&=\lim_{m\to\infty}\P(B_m>b+\eps)=0.
	\end{align*}
The theorem is now proved.
\end{IEEEproof}

In \cref{thm:chisq_est_covp}, condition~\ref{it:g} is implied by the universal approximation property of neural networks \cite{HorTinWhi:J89} for $\bphi$ in a sufficiently high dimensional convex set $\Phi$. The conditions (b)-(d) can be satisfied by choosing proper activation functions for the neural networks.

\subsection{Data-Driven Privacy-Preserving Framework}
\label{sect:dd_frm}
The empirical estimate of the $\chi^2$-divergence empowers us to compute the privacy quantity from data without the need to estimate the distribution of data. In what follows, we employ the $\chi^2$-divergence as a privacy metric and present a data-driven framework for trading off privacy and utility.

A privacy-preserving framework comprises three components: sanitizer, privacy function and utility function. A sanitizer takes the raw data $X$ as input and produces the sanitized data $Y$, in an attempt to remove the statistical information about the private variable $S$ from $X$. In practice, a sanitizer can be realized by a noisy transformation:
\begin{align}\label{eq:san}
	Y=h_{\btheta}(X,N),
\end{align}
where $h_{\btheta}$ is a neural network function parameterized by $\btheta$, and $N$ is the noise perturbation. A naive sanitizer is a constant function, which, however, deprives $Y$ of any utility. It is necessary to reach a compromise between privacy and utility, e.g., requiring that $Y$ is maximally informative about a utility task while not containing an excessive amount of information about $S$.

To learn the optimal sanitizer parameter $\btheta$, we need a privacy function to quantify the information between $S$ and $Y$. In this paper, the square root version of $\chi^2$-divergence \cref{eq:chisq_est} is adopted as the privacy function (taking the square root to counter the vanishing gradient problem). Given a set of samples $\{s_i,x_i\}_{i=1}^m$ drawn from $(S,X)$, we generate $y_i=h_{\btheta}(x_i,n_i)$ (with $n_i$ being a random perturbation) to obtain $\calD_{S,Y}=\{s_i,y_i\}_{i=1}^m$. Then the privacy function $\scP(\btheta;\bphi)$ is formulated as:
\begin{align*}
	\max_{\bphi}\scP(\btheta;\bphi) :=\sqrt{\hat{\chi}^2_m(p_{S,Y}\|\ q_{S,Y})}.
\end{align*}
Note that each $y_i$ is parameterized by the trainable parameter $\btheta$. For a fixed $\btheta$, maximizing $\scP(\btheta;\bphi)$ over $\bphi$ yields an estimate of the dependence between $S$ and $Y$.

On the other hand, a utility function measures the usefulness of the sanitized variable $Y$ w.r.t. a utility variable $U$ of interest. We denote the utility function as $\calL(\btheta;\btau)$, in which $\btau$ is the trainable parameter of the utility model. For example, $\calL(\btheta;\btau)$ can be the reconstruction loss of $X$ from $Y$ by letting $U=X$, and $\btau$ is the vector of model weights. Minimizing $\calL(\btheta;\btau)$ over $\btau$ yields the minimum reconstruction error.

\begin{figure}[htp]
	\centering
	\includegraphics[scale=1]{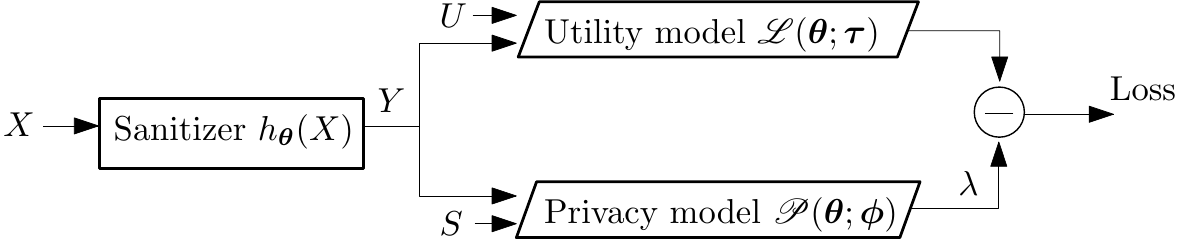}
	\caption{The privacy-preserving framework.}
	\label{fig:put_framework}
\end{figure}

With the privacy and utility functions at hand, optimizing the sanitizer parameter $\btheta$ can be formulated as an unconstrained optimization (\cref{fig:put_framework}):
\begin{align}\label{opt:dd_put}
	\min_{\btheta,\btau}\set*{\calL(\btheta;\btau)
		+\lambda\max\set*{\max_{\bphi}\scP(\btheta;\bphi),\sqrt{\eta}}},
\end{align}
where $\eta$ is the privacy budget for $\chi^2$-divergence privacy and $\lambda$ is a constant to reflect the significance of privacy protection.
The work \cite{WanTay:J20} proposed an alternating algorithm to optimize \cref{opt:dd_put}, which is reproduced in \cref{tab:algo}. Firstly, we freeze $\btheta$ and optimize $\scP(\btheta;\bphi)$ and $\calL(\btheta;\btau)$, respectively. Then, we fix $\btau$ and $\bphi$ and update $\btheta$. These two steps are repeated until an equilibrium is reached.
\begin{algorithm}[!htb]
	\caption{Minibatch stochastic gradient algorithm}\label{tab:algo}
	\begin{algorithmic}[1]
		\STATE{Initialize {$\btheta,\bphi,\btau$}}.
		\REPEAT
		\STATE{Sample a mini-batch set from a training set.}
		\STATE{Optimize $\calL(\btheta;\btau)$ over $\btau$ and optimize $\scP(\btheta;\bphi)$ over $\bphi$.}
		\STATE{Optimize \cref{opt:dd_put} to update $\btheta$.}
		\UNTIL{$\btheta$ converges}
	\end{algorithmic}
\end{algorithm}

It is worth noting that the optimization strategy in \cref{tab:algo} is analogous to the empirical risk approach \cite{Ham:J17,HuaKaiSan:C18}, where finding the optimal sanitization scheme is formulated as a competing game between a sanitizer and an adversary. We demonstrate in \cref{sect:expt} that such approaches are prone to failure as the sanitizer can be fooled by an adversary. Our framework based on $\chi^2$-divergence privacy does not assume that the adversary uses a particular attack model and is thus agnostic to the adversarial attack model. From a theoretical perspective, if the data distribution is known, the $\chi^2$-divergence privacy should be satisfied regardless of the attack that the adversary can muster. Since our framework is data-driven with unknown data distribution, we use the estimate of $\chi^2$-divergence. 

\section{Numerical Experiments}
\label{sect:expt}

In this section, we conduct experiments on the proposed privacy-preserving framework in \cref{sect:dd_frm} to demonstrate the efficacy of the $\chi^2$-divergence privacy metric. 
After training the privacy-preserving framework, we simulate the worst-case privacy attacks (in which the sanitization scheme is known to the attacker). We train an attack model and evaluate the level of privacy protection by the attacker's inference loss of the private variable from the sanitized data.

\subsection{Privacy-Preserving Hypothesis Testing}
\label{sect:synth_expt}

In this experiment, we let $S=\{-1,1\}$ and $U=\{-1,1\}$ be two binary hypotheses, which are statistically dependent on a noisy measurement $X$. The task is to learn the sanitized data $Y$ from $X$ such that the detection error of $U$ is minimized while making it difficult for an unknown attacker to detect $S$ from $Y$.

The noisy measurement is generated as $X=\bA\left[{S'}^2,{U'}^2,S'U',S',U'\right]\T$, where $\bA\in\bbR^{5\times 5}$ is a randomly generated matrix and $S'\sim\N{S}{1}$ and $U'\sim\N{U}{1}$ are noisy observations.

\subsubsection{Network architecture}
The sanitizer function is
$Y=h_{\btheta}(X,N)=X+h'_{\btheta}(N)$,
where $h'_{\btheta}$ is a multilayer perceptron of 5 layers with LeakyRelu activation and $N$ is a Gaussian white noise as a perturbation. The utility function is exactly the loss of a neural classifier \gls{wrt} $U$:
\begin{align*}
	\calL(\btheta;\btau)
	=\sum_{j=1}^m\sum_{i=1}^2f(u_i \mid y_j,\btau)\log{p_{U \mid X}(u_i \mid x_j)}, 
\end{align*}
where $f(u_i \mid y,\btau)$ is the output of the neural classifier, with $\btau$ denoting the trainable parameter and $p(\cdot \mid y)$ denotes the one-hot encoding of the class of input $x_i$, i.e., $p(u_i \mid x_j)=1$ if $x_j$ is labeled with class $u_i$.
The neural classifier is a multilayer perceptron of 5 layers with tanh activation. The generating function $g_{\bphi}$ for the $\chi^2$-divergence privacy metric \cref{eq:chisq_est} is a multilayer perceptron of 5 layers with ELU activation.

We draw $4000$ samples and apply the Adam optimizer with learning rate $10^{-4}$ and batch size $500$ to train the sanitizer according to \cref{tab:algo}.

\subsubsection{Experimental results}
\begin{figure}[htp]
	\centering
	\begin{subfigure}{.49\textwidth}
		\centering
		\includegraphics[width=\textwidth]{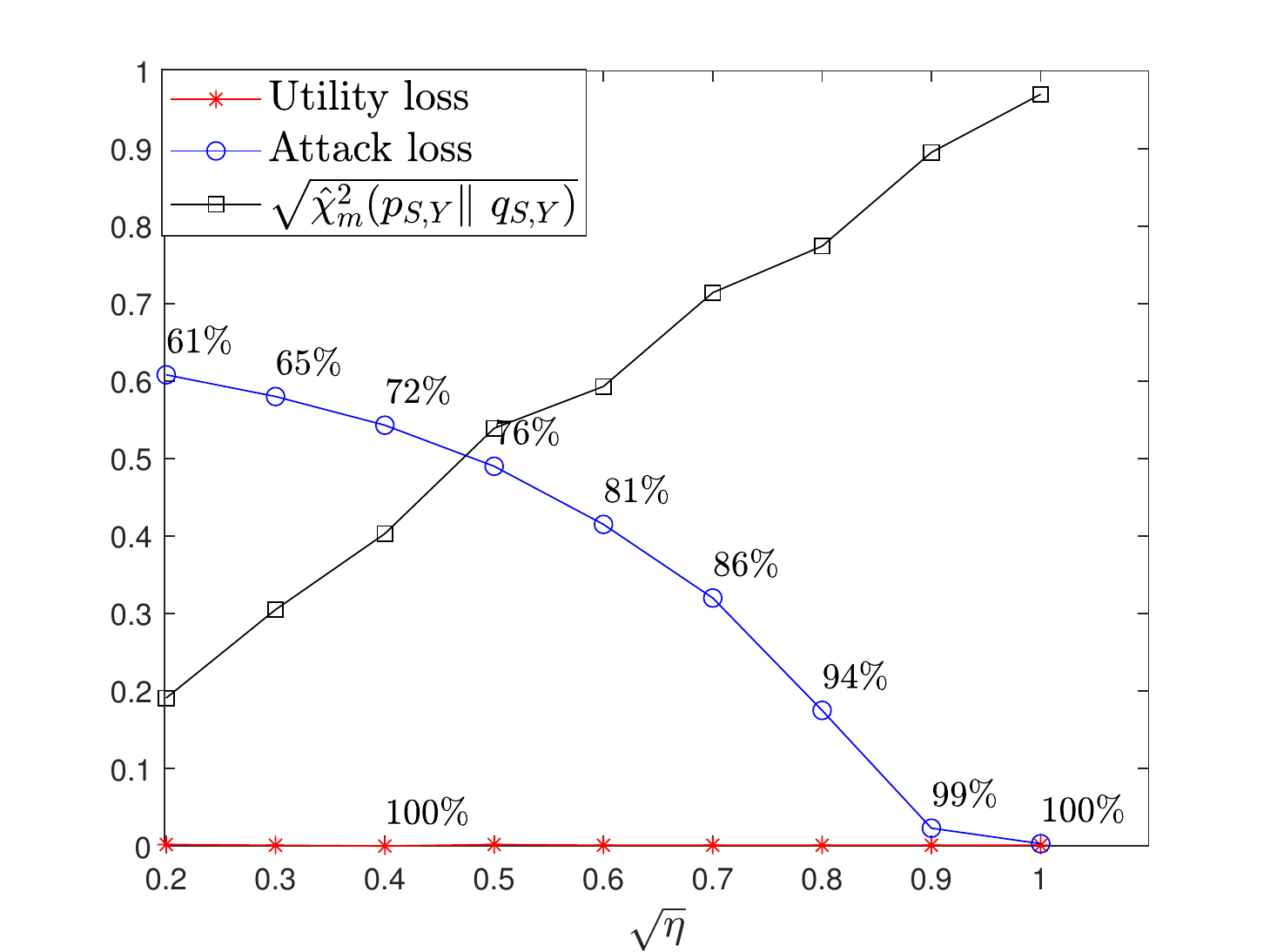}
		\caption{$S$ and $U$ are independent.}
		\label{fig:S_U_not_corr}
	\end{subfigure}
	\begin{subfigure}{.49\textwidth}
		\centering
		\includegraphics[width=\textwidth]{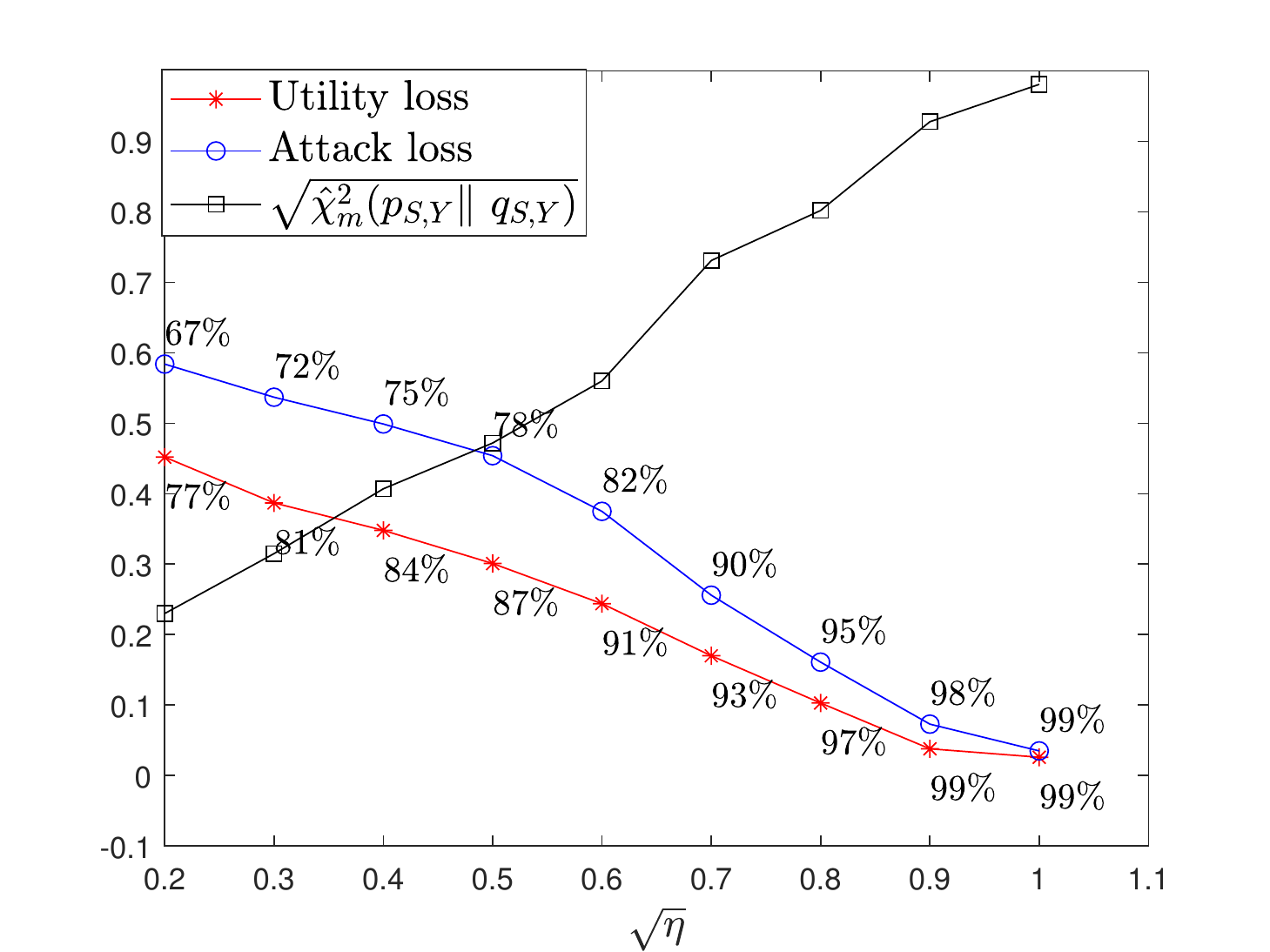}
		\caption{$S$ and $U$ are correlated.}
		\label{fig:S_U_corr}
	\end{subfigure}
	\caption{The privacy-utility trade-offs for hypothesis testing. The percentages shown are classification accuracies.}
\end{figure}

To simulate the privacy attack, we train a neural classifier to detect $S$ from $Y$ after obtaining the sanitizer. We gradually increase the privacy budget $\eta$ and plot the utility loss on $U$ and the attack loss on $S$ (measured in terms of classification accuracy) in \cref{fig:S_U_not_corr,fig:S_U_corr}. In \cref{fig:S_U_not_corr}, $S$ and $U$ are independent with $p_{S,U}(s,u)=0.25$ for each $u$ and $s$. In \cref{fig:S_U_corr}, $S$ and $U$ are correlated with $p_{S,U}(1,1)=p_{S,U}(-1,-1)=0.4$ and $p_{S,U}(-1,1)=p_{S,U}(1,-1)=0.1$. It can be seen that a higher level of privacy protection is at the cost of less utility when $S$ and $U$ are correlated, while the utility is not affected by increasing privacy when $U$ is independent of $S$. A diminishing $\chi^2$-divergence leads to an increasing classification loss on $S$. This suggests that IT privacy metrics can defend against unknown adversarial attacks as alluded to in \cref{sect:itp2pip}. 

\subsection{Privacy-Preserving Auto-Encoders}

In this experiment, we impose the $\chi^2$-divergence privacy metric on variational auto-encoders (VAE) \cite{KinWel:C14} and our task is to learn latent representations of images that are insensitive to a chosen private attribute associated with the images. We compare our method against the generative adversarial privacy (GAP) \cite{HuaKaiSan:C18}, the variational fair autoencoder (VFAE) \cite{ChrKevYuj:C16} and the invariant representation learning (IRL)\cite{DanShuRob:C18} on the UTKface \cite{ZhaSonQi:C17} and CelebA dataset \cite{LiuLuoWan:C15} dataset.

UTKface is a face attribute dataset with annotations of age, gender and ethnicity.
CelebA is a large-scale face attributes dataset with more than 200,000 celebrity images, each with 40 binary attribute annotations. 
We choose the gender attribute as the private variable for UTKface and the smiling attribute as the private variable for CelebA.

\subsubsection{Preliminaries}
Given a high-dimensional input variable $X$, a VAE learns a continuous latent variable $Y$ of the input $X=\bx$ through a reparameterization of the variational lower-bound of $\log p_{X}(\bx)$:
\begin{align}\label{eq:vae}
	\calL(\bx;\btheta,\btau)=\E[\log p_{X\mid Y}(\bx\mid Y)]
	-\KLD{q_{Y \mid X}(\cdot\mid\bx)}{p_Y},
\end{align}
where $q_{Y \mid X}$ is the variational encoder (parameterized by $\btheta$) that approximates the intractable posterior distribution and $p_{X \mid Y}$ is the decoder (parameterized by $\btau$). In this case, the encoder is equivalent to the notion of sanitizer, the utility is the reconstruction loss ($U=X$), and the latent variable $Y$ is the sanitized data. For tractability, it is assumed that $Y\sim\N{\bzero}{\bI}$ and
\begin{align*}
	&q_{Y \mid X}=\N{\bmu(X)}{\diag(\bsigma(X))},\nn
	&p_{X \mid Y}=\N{\bnu(Y)}{\bI},
\end{align*}
in which $\bmu(\cdot)$ and $\bsigma(\cdot)$ are neural network functions with their collective trainable weights denoted by $\btheta$. The function $\bnu(\cdot)$ is a neural network function with trainable weights denoted by $\btau$. Given a training set $\set{\bx_i}_{i=1}^m$, the utility function can be written as 
\begin{align*}
	\calL(\btheta;\btau)
	=-\sum_{i=1}^m{\calL(\bx_i;\btheta,\btau)},
\end{align*}
which is to be minimized over $\btheta$ and $\btau$. Following the framework \cref{opt:dd_put}, the $\chi^2$-divergence privacy metric is used for encouraging the disentanglement of $S$ and $Y$.

The original VAE serves as the baseline. The GAP framework differs from our $\chi^2$-divergence method \cref{opt:dd_put} in that GAP quantifies privacy using the empirical risk of an adversary model \cite{HuaKaiSan:C18} instead of an agnostic privacy function. The VFAE and IRL, which are variants of VAEs, aim to factor out a sensitive variation from the latent variable and are thus on a comparable basis with our method. In contrast to our method and GAP, the encoders of the VFAE and IRL (i.e., $p_{Y \mid S,X}(Y \mid S,X)$) take an additional input of the private attribute. Therefore, the sanitizer (i.e., the encoder) needs to know the label of $S$ for $X$. To penalize privacy leakage, the VFAE uses the maximum mean discrepancy between $p_{Y \mid S}(\cdot \mid s_i)$ and $p_{Y \mid S}(\cdot \mid s_j)$ for $s_i\neq s_j$, while the IRL uses the pairwise KL divergences $\KLD{p_{Y \mid S, X}(\cdot \mid s_i,\bx_i)}{p_{Y \mid S,X}(\cdot \mid s_j,\bx_j)}$ for $i\neq j$. For detailed VFAE and IRL frameworks, we refer readers to \cite{ChrKevYuj:C16} and \cite{DanShuRob:C18}, respectively. The privacy function is multiplied by a constant $\lambda$ (similar to $\lambda$ in \cref{opt:dd_put}).

\subsubsection{Experimental Setup}
The VAE encoder networks $\bmu(\cdot)$ and $\log\bsigma(\cdot)$ share $6$ down-sampling ResNet blocks \cite{HeZhaRena:C16} followed by two separate dense layers. The VAE decoder network $\bnu(\cdot)$ is made of a dense layer and $6$ up-sampling convolutional layers that recover the input image size. The dimension of the latent variable $Y$ is 4608. This network architecture also applies to VFAE and IRL except that an additional channel for feeding $S$ is required at the input of the encoder and decoder. 
The generating function $g_{\bphi}$ for the $\chi^2$-divergence is made of $4$ MLPs with hidden units (2304, 1152, 576, 1) with Instance Normalization.
The adversary (for GAP) and attack models (for evaluating privacy leakage) are MLPs of $4$ layers with hidden units (2304, 1152, 576, 1).

For training, we use the Adam optimizer with $10^{-4}$ learning rate and $0.5$ (reps. $0.99$) momentum for running average mean and (resp. square).

\subsubsection{Experimental Results}
\begin{figure}[!htp]
	\centering
	\includegraphics[scale=0.55]{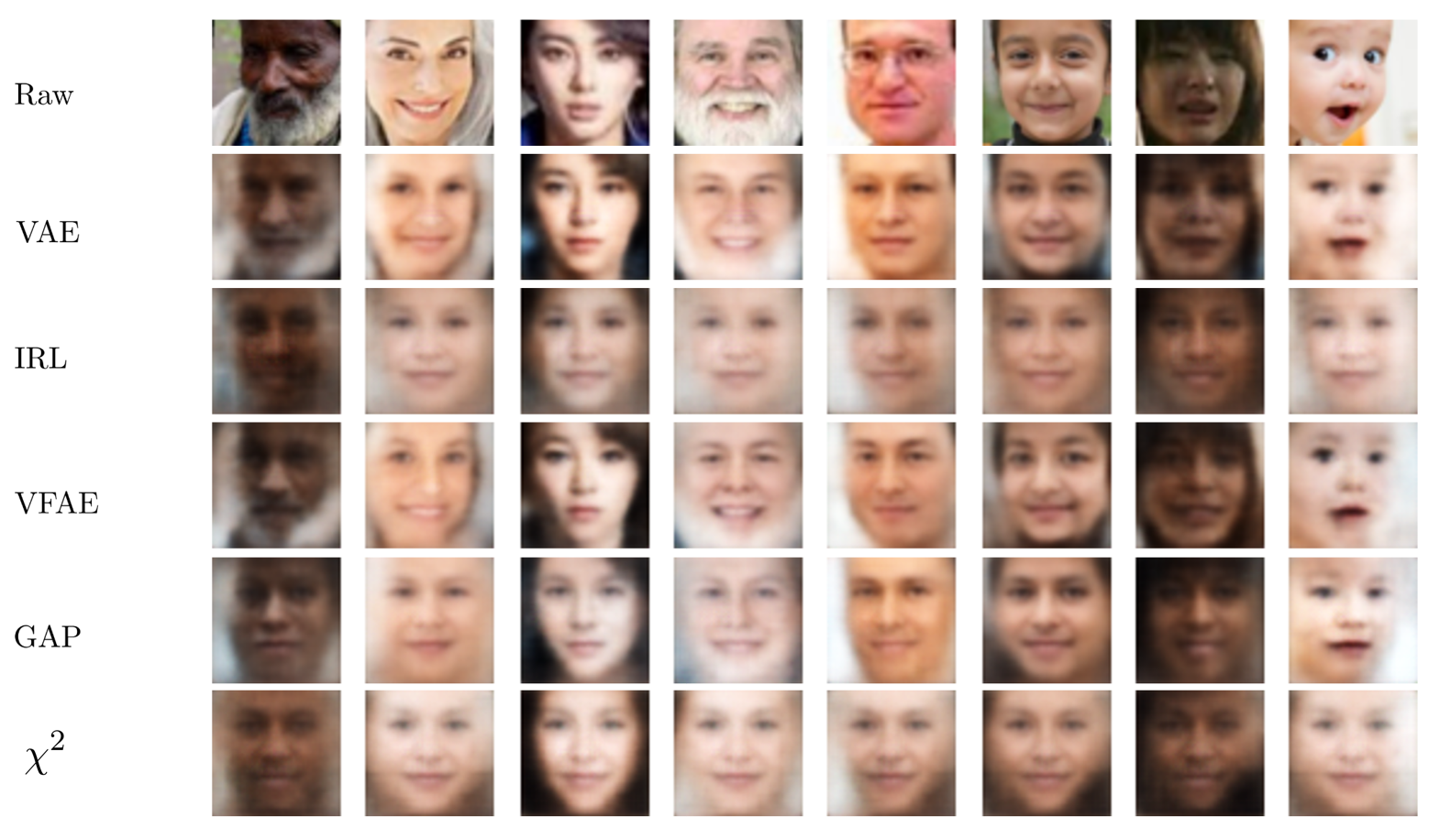}
	\caption{Reconstructed UTKface images from the latent space where gender is the private attribute.}
	\label{fig:sanitized_utk}
\end{figure}

The mean square error (MSE) for reconstruction and the attack loss and accuracy for UTKface are shown in \cref{tab:results_utk}.\footnote{Abbreviations. Prv.: Private, Attr.: Attribute, Acc.: Accuracy, Util.: Utility.} 
A $\downarrow$ symbol means a smaller value is better and vice versa for the $\uparrow$ symbol.
Samples of the reconstructed images are displayed in \cref{fig:sanitized_utk}. We set $\lambda=20$ for GAP and choose $\sqrt{\eta}=0.1$ and $\lambda=5$ for our method so that it has an attack performance similar to that of GAP. From the reconstruction MSE, it can be seen that GAP and our method generate a similar utility loss. However, the adversary model in GAP is identical to the attack model. If we replace the batch normalization with instance normalization for the adversary model in GAP (whose results are shown in GAP-A), the level of privacy protection dropped significantly as indicated by the attack performance. Therefore, privacy cannot be ensured by the empirical risk if the adversary model in GAP does not match the attack model.

\begin{figure}[!htp]
	\centering
	\includegraphics[scale=0.6]{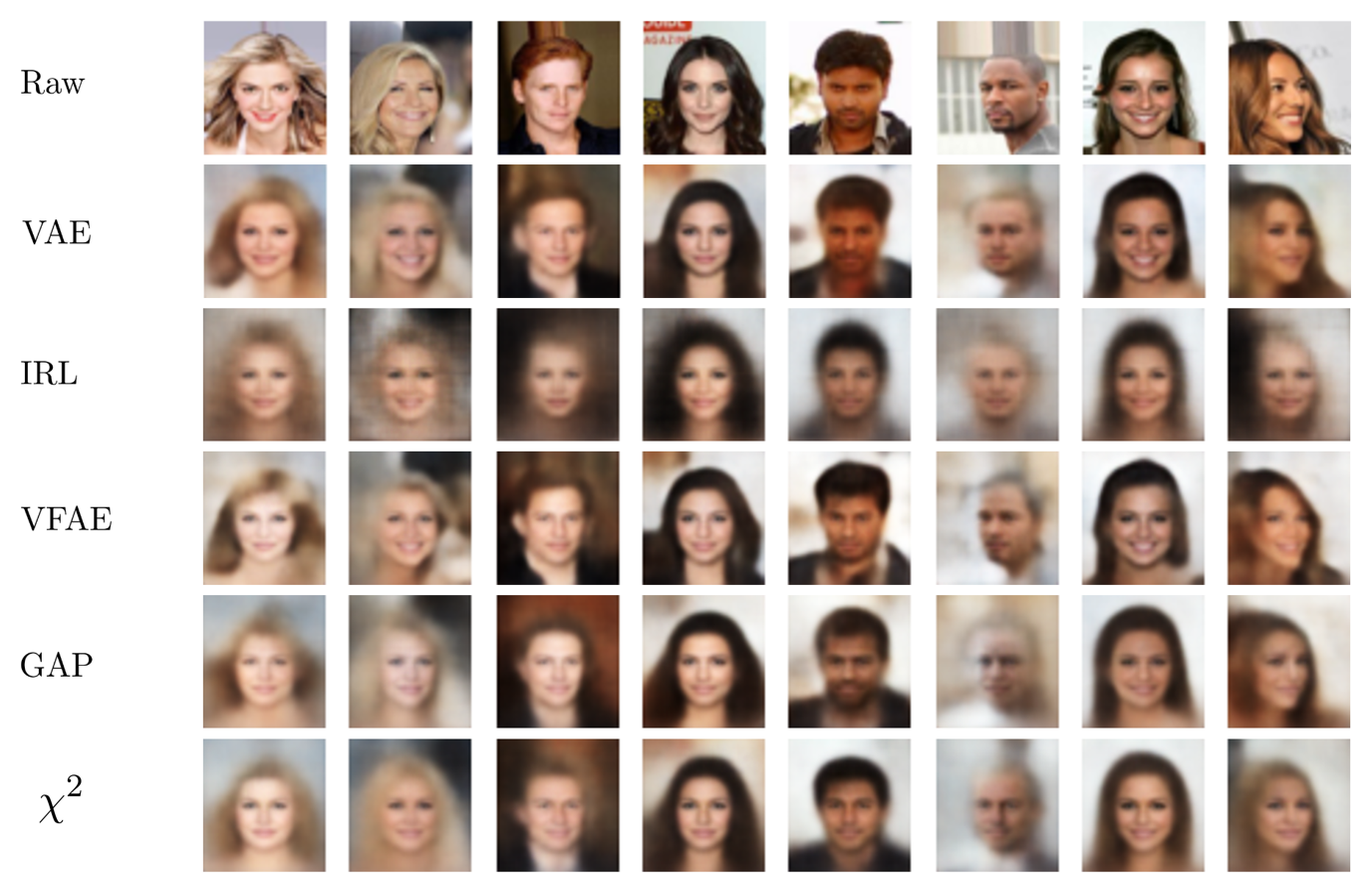}
	\caption{Reconstructed Celebfaces images from the latent space where smiling is the private attribute and gender classification is the utility task.}
	\label{fig:sanitized_celeb}
\end{figure}

The results for CelebA are shown in \cref{tab:results_celeb} with samples of reconstructed images displayed in \cref{fig:sanitized_celeb}. In this case, we include an additional utility task of classifying gender in the learning architecture. Retaining the $\lambda$ and $\eta$ used for UTKface, our method outperforms the GAP (where the adversary model and attack model are the same) in terms of privacy protection. Adversarial training is known to be unstable and the quality of privacy sanitization is determined by the capability of the chosen adversarial neural network, which in practice cannot incorporate all possible adversarial strategies. In contrast, the $\chi^2$-divergence privacy metric captures statistical information from data without assuming an adversary model.

In both cases, VFAE failed to remove the private attributes while severely distorting the data (leading to a large reconstruction error). We made attempts to improve the VFAE performance by changing $\lambda$. However, the privacy protection offered by the VFAE is not controllable by $\lambda$. IRL with $\lambda=50$ achieves its best privacy protection across different values of $\lambda$ but is still weaker than our method and GAP. The accuracy of classifying the utility variable is better preserved for our method when compared to the VAE baseline. Results in \cref{sect:synth_expt} suggest that a utility variable can be preserved almost intact if it is independent of the private variable.

\begin{table*}[!htb]
	\centering
	\caption{UTKface dataset.}
	\label{tab:results_utk}
	\begin{tabular}{lccccccc}
		\toprule
			                         & VAE   & VFAE & IRL   & GAP   & GAP-A & \textbf{$\chi^2$} \\
		\midrule
		Prv. Attr. Acc. $\downarrow$ & 88\%  & 98\% & 84\%  & 70\%  & 83\%  & 69\% \\
		Prv. Attr. Loss $\uparrow$   & 0.29  & 0.07 & 0.37  & 0.56  & 0.37  & 0.58 \\
		Util. MSE       $\downarrow$ & 0.026 & 0.08 & 0.029 & 0.057 & 0.041 & 0.07 \\
		\bottomrule
	\end{tabular}
\end{table*}
\begin{table*}[!htb]
	\centering
	\caption{CelebFaces dataset.}
	\label{tab:results_celeb}
	\begin{tabular}{ccccccc}
		\toprule
		                              & VAE    & VFAE   & IRL    & GAP    & GAP-A  & \textbf{$\chi^2$} \\
		\midrule
		Prv. Attr. Acc.  $\downarrow$ & 85\%   & 99.5\% & 75\%   & 79\%   & 82\%   & 66\% \\
		Prv. Attr. Loss  $\uparrow$   & 0.36   & 0.015  & 0.48   & 0.45   & 0.41   & 0.61 \\
		Util. MSE        $\downarrow$ & 0.04   & 0.12   & 0.036  & 0.06   & 0.05   & 0.075\\
		Util. Attr. Acc. $\uparrow$   & 99.7\% & 93\%   & 98\%   & 98.8\% & 99\%   & 98\% \\
		Util. Attr. Loss $\downarrow$ & 0.006  & 0.17   & 0.057  & 0.035  & 0.03   & 0.06 \\
		\bottomrule
	\end{tabular}
\end{table*}

\section{Conclusion} \label{sect:conclusion}
In this paper, we have made connections between probabilistic IP and weak DP and shown that imposing this privacy notion leads to error lower bounds for detecting and estimating the private variable from the sanitized variable. 
Based on probabilistic IP, we characterized several well-known IT privacy metrics given by $f$-divergences. We argued that $\chi^2$-divergence privacy is stronger than TV and KL divergence privacy metrics. Therefore, we used $\chi^2$-divergence to develop a data-driven privacy-preserving framework. In this paper, we have not investigated the analytical bounds for privacy-utility trade-offs under $\chi^2$-divergence privacy. An interesting future work is to consider different utility measures and derive fundamental trade-off bounds if they exist.

\appendices

\def\appa{\ref{it:TV}}

\section{Proof of \cref{thm:itp2pip}{\normalfont \ref{it:TV}}}
\label[Appendix]{proof:itp2pip_a}

Since $\calL_{\eps}\cap\calR_{\eps}=\emptyset$, we have
\begin{align*}
	\TV{p_{S,Y}}{q_{S,Y}}
	&=\int_{\calS\times\calY}\abs*{p_{S,Y}(s,y)-p_S(s)p_Y(y)}\ud{s}\ud{y}\nn
	&\geq\int_{\calL_{\eps}\cup\calR_{\eps}}\abs*{p_{S,Y}(s,y)-p_S(s)p_Y(y)}\ud{s}\ud{y}\nn
	&\geq (\exp^{\eps}-1)\int_{\calL_{\eps}}p_{S,Y}(s,y)\ud{s}\ud{y}
	+(1-\exp^{-\eps})\int_{\calR_{\eps}}p_{S,Y}(s,y)\ud{s}\ud{y}\nn
	&=(\exp^{\eps}-1)\P(\calL_{\eps})+(1-\exp^{-\eps})\P(\calR_{\eps})\nn
	&\geq(1-\exp^{-\eps})\P(\calL_{\eps})+(1-\exp^{-\eps})\P(\calR_{\eps})\nn
	&=(1-\exp^{-\eps})\P(\calL_{\eps}\cup\calR_{\eps}),
\end{align*}
where the last inequality is due to $\exp^{\eps}-1\geq 1-\exp^{-\eps}$. Finally, we have
\begin{align*}
	\TV{p_{S,Y}}{q_{S,Y}}\leq\eta
	\implies\P(\calL_{\eps}\cup\calR_{\eps})\leq\frac{\eta}{1-\exp^{-\eps}},
\end{align*}
and the proof is complete.

\section{Proof of \cref{thm:itp2pip}{\normalfont \ref{it:KL}}}
\label[Appendix]{proof:itp2pip_b}
For an arbitrary event $A\in\calF$, consider a channel that produces a Bernoulli random variable $W$ based on the following law:
$p_{W\mid S,Y}(1\mid s,y)=1$ if $S^{-1}(s)\cap Y^{-1}(y)\cap A \ne\emptyset$ and $0$ otherwise.
Then the distribution of $W$, when $(S,Y)$ is generated by $p_{S,Y}$, is $p_W(1)=p$, where
\begin{align*}
	p=\int_{A}p_{S,Y}(s,y)\ud{s}\ud{y}.
\end{align*}
And the distribution of $W$, when $(S,Y)$ is generated by $q_{S,Y}$, is $q_W(1)=q$, where
\begin{align*}
	q=\int_{A}q_{S,Y}(s,y)\ud{s}\ud{y}.
\end{align*}
From the data processing inequality, we have
\begin{align}
	\KLD{p_{S,Y}}{q_{S,Y}}
	&\geq\KLD{p_W}{q_W}\nn
	&=p\log\frac{p}{q}+(1-p)\log\frac{1-p}{1-q}. \label{bit_kld}
\end{align}
Let $\gamma=\dfrac{p}{q}$ and the right-hand side of \cref{bit_kld} can be written as
\begin{align*}
	f(p,\gamma)=\log\gamma+(1-p)\log\frac{1-p}{\gamma-p}.
\end{align*}
The partial derivatives of $f(p,\gamma)$ are
\begin{align*}
	&\ppfrac{f(p,\gamma)}{p}=\frac{1-\gamma}{\gamma-p}-\log \frac{1-p}{\gamma-p} \geq 0,\nn
	&\ppfrac{f(p,\gamma)}{\gamma}=\frac{p(\gamma-1)}{\gamma(\gamma-p)} 
	 \left\{\begin{array}{ll}
	\leq 0 &\ \text{if}\ \gamma < 1,\\
	\geq 0 &\ \text{otherwise},
	\end{array}\right.
\end{align*}
where the inequalities are due to $\gamma=\dfrac{p}{q}>p$. Therefore, it can be concluded that 
\begin{itemize}
	\item For any fixed $\gamma>0$, $f(p,\gamma)$ is non-decreasing \gls{wrt} $p \in [0,1]$.
	\item For any fixed $p \in [0,1]$, $f(p,\gamma)$ is non-increasing \gls{wrt} $\gamma<1$, and non-decreasing \gls{wrt} $\gamma\geq1$.
\end{itemize}

Now letting $A=\calL_{\eps}$, we have $\gamma\leq\exp^{-\eps}$ and $p=\P(\calL_{\eps})$. From the claim assumption and \cref{bit_kld}, we have $f(p,\gamma)\leq\eta$. Consequently, we obtain
\begin{align*}
	\P(\calL_{\eps})=p
	\leq\sup\set*{p'\in\left[0,1\right]\given f(p',\exp^{-\eps})\leq\eta}.
\end{align*}

On the other hand, letting $A=\calR_{\eps}$, we have $\gamma\geq\exp^{\eps}$ and $p=\P(\calR_{\eps})$. Similarly, we must have
\begin{align*}
	\P(\calR_{\eps})=p
	\leq\sup\set*{p'\in\left[0,1\right]\given f(p',\exp^{\eps})\leq\eta}.
\end{align*}
The proof is completed by noting that
$\P(\calL_{\eps}\cup\calR_{\eps})
=\P(\calL_{\eps})+\P(\calR_{\eps})$.

\section{Proof of \cref{thm:itp2pip}{\normalfont \ref{it:Chi}}}
\begin{figure}[!htb]
	\centering
	\includegraphics[scale=0.5]{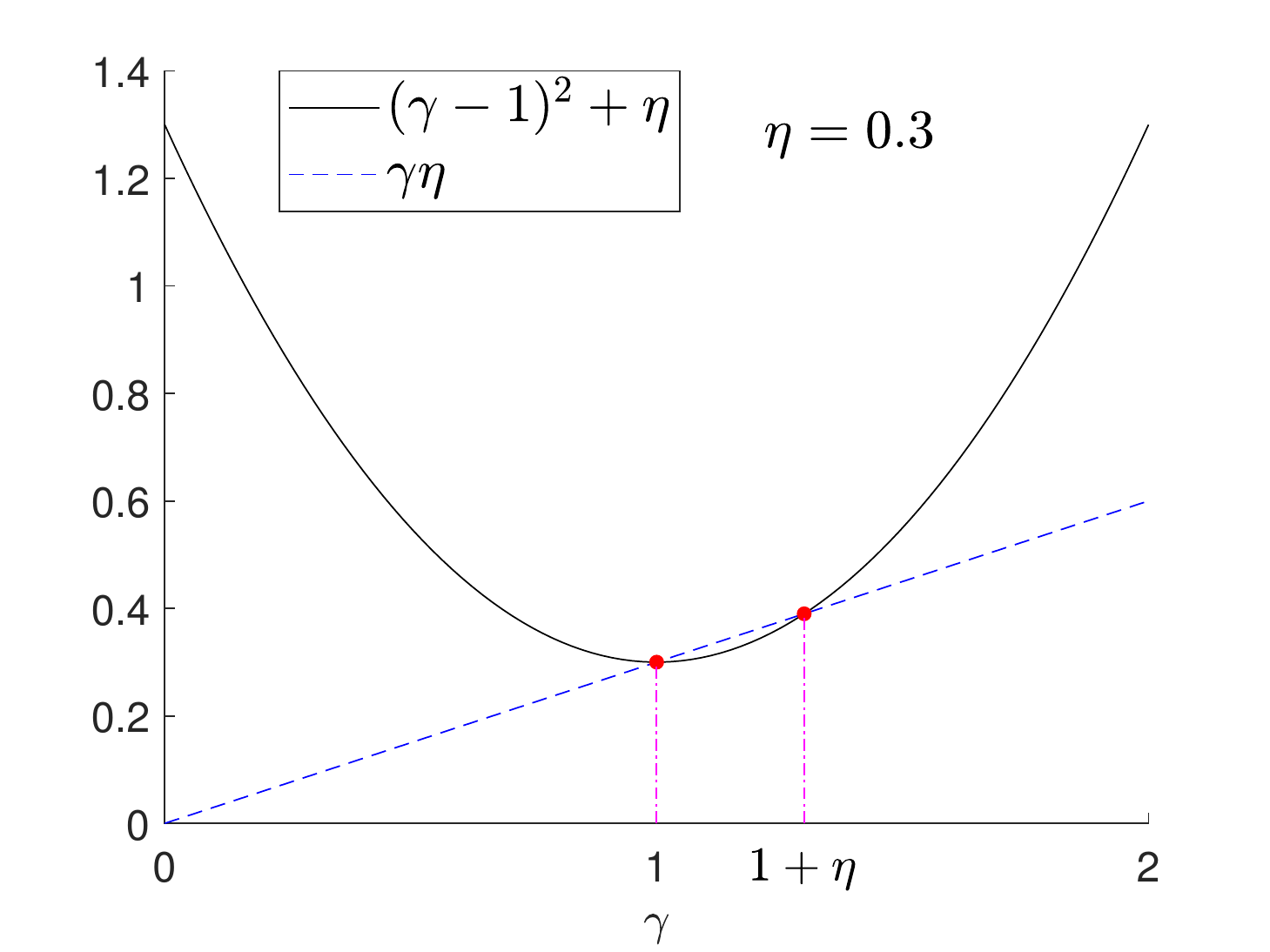}
	\caption{The denominator and numerator of $g(\gamma,\eta)$.}
	\label{fig:chisq_delta}
\end{figure}

\label[Appendix]{proof:itp2pip_c}
The proof exploits the geometric property of $\chi^2$-divergence. 
Let $A\in\calF$ be an arbitrary event. From Sedrakyan's inequality (which is a direct consequence of the  Cauchy-Schwarz inequality), we have
\begin{align}
	\ChiSq{p_{S,Y}}{q_{S,Y}}
	&=\int_{A\cup A\setcomp}\frac{p_{S,Y}(s,y)^2}{p_S(s)p_Y(y)}\ud{s}\ud{y}-1\nn
	&\geq\frac{p^2}{q}+\frac{(1-p)^2}{1-q}-1,\label{chi_sq_ineq}
\end{align}
where 
\begin{align*}
	&p=\int_{A}p_{S,Y}(s,y)\ud{s}\ud{y}=\P(A),\nn
	&q=\int_{A}p_S(s)p_Y(y)\ud{s}\ud{y}.
\end{align*}
Let $\gamma=\dfrac{p}{q}$. Substituting $q=\dfrac{p}{\gamma}$ into \cref{chi_sq_ineq} and from the assumption $\ChiSq{p_{S,Y}}{q_{S,Y}}\leq\eta$, we obtain
\begin{align*}
	p\gamma
	+\frac{\left(1-p\right)^2}{1-p/\gamma}-1
	\leq\eta.
\end{align*}
Rearranging the above inequality, we have  
\begin{align}\label{chi_sq_tail}
	\P(A) = p \leq g(\gamma,\eta)
\end{align}
where
\begin{align*}
	g(\gamma,\eta)=\frac{\gamma\eta}{(\gamma-1)^2+\eta}.
\end{align*}
The following properties about $g(\gamma,\eta)$ can be verified by checking its derivatives. (For the reader's convenience, we visualize $g(\gamma,\eta)$ by plotting its numerator and denominator as functions of $\gamma$ in \cref{fig:chisq_delta}.)
\begin{itemize}
	\item For a fixed $\eta>0$, $g(\gamma,\eta)$ is monotonically increasing \gls{wrt} $\gamma^2\in\left[0,1+\eta\right]$ and monotonically decreasing \gls{wrt} $\gamma^2\in\left(1+\eta,\infty\right)$.
	\item $g(\gamma,\eta)\geq 1$ for $\gamma\in\left[1,1+\eta\right]$.
\end{itemize}
	
Now we substitute $\calL_{\eps}$ and $\calR_{\eps}$ for $A$ in \cref{chi_sq_tail}. It can be verified that $\gamma\leq\exp^{-\eps}$ when $A=\calL_{\eps}$, and $\gamma\geq\exp^{\eps}$ when $A=\calR_{\eps}$.
From the monotonicity property of $g(\gamma,\eta)$, we have
\begin{align*}
	&\P(\calL_{\eps})\leq\frac{\exp^{-\eps}\eta}{(\exp^{-\eps}-1)^2+\eta},\ \forall \eps>0,\nn
	&\P(\calR_{\eps})\leq\frac{\exp^{\eps}\eta}{(\exp^{\eps}-1)^2+\eta},\ \forall \eps>\log(1+\eta).
\end{align*}
Note that the second inequality above also holds true for $\eps\in\left[0,\log(1+\eta)\right]$ because $\P(\calR_{\eps})\leq 1$ while its right-hand side is greater than $1$. The proof for \cref{it:Chi} is now complete.

\section{Proof of \cref{lemma:tv2pip}}
\label[Appendix]{proof:tv2pip}
Let 
\begin{align*}
	\gamma(A)=\int_{A} (p_{S,Y}(s,y)-p_S(s)p_Y(y))\ud{s}\ud{y},
\end{align*}
and denote 
\begin{align*}
	&\Psi
	=\set*{\omega\given\exp^{-\eps}\leq d(S(\omega),Y(\omega))\leq1},\nn
	&\Gamma
	=\set*{\omega\given 1\leq d(S(\omega),Y(\omega))\leq\exp^{\eps}}.
\end{align*}
Firstly, we have
\begin{align}
	\gamma(\Gamma)-\gamma(\Psi)
	&\leq(1-\exp^{-\eps})\P(\Gamma)+(\exp^{\eps}-1)\P(\Psi)\nn
	&\leq(\exp^{\eps}-1)\left(\P(\Gamma)+\P(\Psi)\right)\nn
	&\leq\exp^{\eps}-1. \label{tv2pip:ineq}
\end{align}
Moreover, we have
\begin{align*}
	\gamma(\calR_{\eps})
	\leq\int_{\calR_{\eps}}p_{S,Y}\ud{s}\ud{y}
	\leq\delta.
\end{align*}
From $\gamma(\calL_{\eps})+\gamma(\calR_{\eps})+\gamma(\Gamma)+\gamma(\Psi)=\gamma(\Omega)=0$ and \cref{tv2pip:ineq}, we obtain
\begin{align*}
	-\gamma(\calL_{\eps})
	&=\gamma(\Gamma)+\gamma(\Psi)+\gamma(\calR_{\eps})\nn
	&\leq\gamma(\Gamma)-\gamma(\Psi)+\delta\nn
	&\leq\exp^{\eps}-1+\delta.
\end{align*}
Finally, we obtain
\begin{align*}
	\TV{p_{S,Y}}{q_{S,Y}}
	&=\gamma(\Gamma)-\gamma(\Psi)+\gamma(\calR_{\eps})-\gamma(\calL_{\eps})\nn
	&\leq 2(\exp^{\eps}-1+\delta),
\end{align*}
and the proof is complete.

\section{Proof of \cref{lemma:max_corr}}\label[Appendix]{proof:lemma:max_corr}

Let $L^2(p_S)$ (resp. $L^2(p_Y)$) be the space of all real-valued functions of $S$ (resp. $Y$) with finite variance. Define a linear operator $T:L^2(p_Y)\to L^2(p_S)$ such that for $f\in L^2(p_Y)$,
\begin{align*}
	[Tf](s)=\E[f(Y) \mid S=s].
\end{align*}
It is associated with an adjoint operator $[T^*g](y)=\E[g(S) \given Y=y]$ for $g\in L^2(p_S)$. Let $(\sigma_i)_{i\geq 1}$ be a sequence of singular values of the operator $T$ in descending order. From the definition of maximal correlation, it is well-known that $\sigma_1=1$ and $\sigma_2=\rho_m(S,Y)$ \cite{Ren:J59}. Moreover, we have 
\begin{align}\label{eq:T_norm_a}
	\norm{T}_{\mathrm{HS}}^2=\sum_{i\geq 1}\sigma_i^2,
\end{align}
where $\norm{\cdot}_{\mathrm{HS}}^2$ is the Hilbert-Schmidt norm.

On the other hand, we can rewrite $T$ as
\begin{align*}
	[Tf](s)=\int_{\calY}f(y)k(s,y)p_Y(y)\ud{y},
\end{align*}
in which $k(s,y):\calS\times\calY\to\bbR$ is a kernel:
\begin{align*}
	k(s,y)=\frac{p_{S,Y}(s,y)}{p_S(s)p_Y(y)}.
\end{align*}
From \cite[Lemma 4.8]{Con:B90}, we have 
\begin{align}
	\norm{T}_{\mathrm{HS}}^2
	&=\int_{\calY}\int_{\calS}k(s,y)^2p_S(s)p_Y(y)\ud{s}\ud{y}\nn
	&=\ChiSq{p_{S,Y}}{q_{S,Y}}+1.\label{eq:T_norm_b}
\end{align}
The proof is completed by combining \cref{eq:T_norm_a,eq:T_norm_b}.

\section{Proof of \cref{lemma:bounded_seq}}
\label[Appendix]{proof:bounded_seq}
Let $v_m(\bphi)$ be the numerator of $\gamma_m(\bphi)$ and
$d_m(\bphi)=\ofrac{m}\sum_{i=1}^mg_{\bphi}(z_i)^2$.
The gradient of $\gamma_m(\bphi)$ can then be written as
$\nabla\gamma_m(\bphi)=\dfrac{k_m(\bphi)}{\left(d_m(\bphi)+\lambda_m\right)^2}$ with
\begin{align*}
	k_m(\bphi)=d_m(\bphi)\nabla v_m(\bphi)-v_m(\bphi)\nabla d_m(\bphi).
\end{align*}

By assumption, $g_{\bphi}(x)$ is a smooth function \gls{wrt} $\bphi$ and continuous \gls{wrt} $x$. Therefore, $\ppfrac{g_{\bphi}(x)}{\bphi}$ is also a continuous function, which is thus uniformly bounded by some constant due to the compactness of $\Phi$ and $\calY$. Therefore, for any $m>0$, $k_m(\bphi)$ (consisting of the mean of bounded functions) is bounded by a constant vector $c\bone$ with $c<\infty$.
Let
\begin{align*}
	B_m=\dfrac{c}{(\min_{\bphi\in\Phi}d_m(\bphi)+\lambda_m)^2},
\end{align*}
which yields $\nabla\gamma_m(\bphi)\leq B_m\bone$ followed by
\begin{align*}
	\abs{\gamma_m(\bphi)-\gamma_m(\bphi')}
	&\leq B_m\bone\T\abs{\bphi-\bphi'}\nn
	&\leq B_m\norm{\bphi-\bphi'}.
\end{align*}

From the uniform law of large numbers \cite{Jen:J69}, we have
\begin{align*}
	\min_{\bphi\in\Phi}d_m(\bphi)
	\convas 
	a\coloneqq\min_{\bphi\in\Phi}\E[d_m(\bphi)],
\end{align*}
where $a>0$. 
As a result, we have 
\begin{align*}
	\lim_{m\to\infty}\P(B_m>\frac{c}{a^2})
	=\lim_{m\to\infty}\P(\min_{\bphi\in\Phi}d_m(\bphi)<a-\lambda_m)=0.
\end{align*}
The proof is now complete.


\bibliographystyle{IEEEtran}
\bibliography{IEEEabrv,StringDefinitions,BibBooks,refs}

\end{document}